\documentclass[pdflatex,sn-mathphys-num]{sn-jnl}

\usepackage{graphicx}%
\usepackage{multirow}%
\usepackage{amsmath,amssymb,amsfonts}%
\usepackage{mathrsfs}%
\usepackage[title]{appendix}%
\usepackage{xcolor}%
\usepackage{textcomp}%
\usepackage{manyfoot}%
\usepackage{orcidlink}%
\usepackage{booktabs}%
\graphicspath{{figures/}}%
\usepackage{tikz}%
\usetikzlibrary{positioning,arrows.meta,fit,calc}%

\definecolor{fsmteal}{HTML}{00687F}
\definecolor{fsmred}{HTML}{EE3342}
\definecolor{fsmdeep}{HTML}{0B3C49}
\definecolor{fsmsky}{HTML}{6FB3C6}
\definecolor{fsmslate}{HTML}{6C757D}

\begin{document}

\title[Do Layer Rankings Survive Silence Trimming?]{Do Layer Rankings
Survive Silence Trimming? The Silence Shortcut in Frozen WavLM
Representations for Audio Deepfake Detection}

\author[1]{\fnm{Nur} \sur{Elabid}\,\orcidlink{0009-0003-2620-989X}}\email{nur.elabid@stu.fsm.edu.tr}
 
\author*[2]{\fnm{Mahmud Esad} \sur{Arar}\,\orcidlink{0000-0002-7472-4189}}\email{mearar@fsm.edu.tr}

\affil[1]{\orgdiv{Department of Biomedical Engineering},
  \orgname{Fatih Sultan Mehmet Vakif University},
  \city{Istanbul}, \country{T\"urkiye}}
 
\affil[2]{\orgdiv{Department of Electrical and Electronics Engineering},
  \orgname{Fatih Sultan Mehmet Vakif University},
  \city{Istanbul}, \country{T\"urkiye}}

\abstract{Layer-wise analyses of frozen self-supervised encoders are used to
locate spoofing information in audio deepfake detection. On the ASVspoof 2019
logical access (LA) benchmark, bonafide recordings carry longer leading and
trailing silence than spoofed ones, and detectors exploit this cue, the silence
shortcut. This paper examines whether layer rankings for a frozen WavLM Base
encoder survive the trimming of this silence. A small classifier is trained on
each of the 12 layers with and without trimming and evaluated on ASVspoof 2019
LA and, zero-shot, on In-the-Wild, in all four train/test combinations, with
speaker-level bootstrap intervals. Trimming at 40~dB raises the in-domain error
at every layer and preserves the layer ranking only partly (Kendall's $\tau$ of
0.58 in domain, 0.73--0.76 on In-the-Wild). The best layer is unchanged at
40~dB, but on the evaluation set it moves from layer~5 to layer~6 at 15~dB
and, with mean pooling, at 20~dB, thresholds at which trimming may also
remove low-energy speech. At 40~dB, the equal error rate of a model
trained with silence rises by 6.3--19.9 percentage points at every layer when
only its test audio is trimmed. On In-the-Wild, the effect depends on the layer: at
layers 3--5, trimming the test audio lowers the error only for models trained
on trimmed audio, whereas at layers 6, 8, and 9 training on trimmed audio
lowers it also on clean test audio. We quantify how zero padding alters WavLM
features through group normalization, an effect that an attention mask does
not fully undo.}


\maketitle

\section{Introduction}\label{sec:intro}

Speech synthesis and voice conversion systems~\cite{tan2021survey} have
advanced to the point where synthetic speech is often indistinguishable from
genuine recordings by human listeners. This capability has legitimate
applications in accessibility, entertainment, and human--computer interaction,
but it also enables impersonation attacks against voice-based authentication,
fraudulent audio evidence, and large-scale disinformation~\cite{yi2023audio}.
Automatic speaker verification (ASV) systems are particularly exposed, since
they were designed to reject impostors rather than synthetic replicas of
enrolled speakers~\cite{kinnunen2020tandem}. Spoofing countermeasures (CMs) are
therefore a practical requirement, and the ASVspoof challenge series has
established standardized benchmarks against which they are
evaluated~\cite{wang2020asvspoof}.

Most recent CMs use a large self-supervised learning (SSL) speech model, such
as wav2vec~2.0~\cite{baevski2020wav2vec}, HuBERT~\cite{hsu2021hubert},
WavLM~\cite{chen2022wavlm}, or XLS-R~\cite{babu2022xlsr}, as a front end,
usually fine-tuned jointly with a detection
head~\cite{wang2022investigating,guo2024audio}. Joint optimization makes it
difficult to separate what the pretrained representations already encode from
what fine-tuning adds~\cite{elkheir2025comprehensive}. Keeping the encoder
frozen removes this confound: the representation is fixed, a single forward
pass yields every layer, and differences between configurations can be
attributed to the downstream choice under test. Layer-wise analysis of frozen
or partially adapted encoders has accordingly become a common way of asking
where spoofing-relevant information resides
\cite{elkheir2025comprehensive,pan2024attentive,sinha2024improving,beheshti2026probing,serrano2025}.
For WavLM Base, intermediate layers have been reported as the most informative,
with layer~4 the best single layer in a recent comparison of six frozen SSL
models~\cite{serrano2025}.

These conclusions rest on benchmarks with a known shortcut. In the ASVspoof
2019 logical access (LA) database, bonafide recordings carry longer leading and
trailing silence than spoofed ones, and detectors exploit this cue. M\"uller
et al.~\cite{muller2021speech} showed that silence duration alone separates the
two classes well above chance and that removing silence raises the error of a
trained detector severalfold. Zhang et al.~\cite{zhang2023impact} analyzed the
effect across attack types, and the issue has been discussed in the ASVspoof
2021 overview~\cite{liu2023asvspoof2021} and in the wider context of shortcut
learning in spoofing detection~\cite{shim2023}. If a layer ranks highly partly
because it encodes silence well, a layer ranking measured on this benchmark is
partly a ranking of how well each layer reads silence. Two studies come close
to testing this. Wang et al.~\cite{wang2024silence} trained and tested ten CMs,
including fine-tuned wav2vec~2.0 systems, on audio with and without silence,
covering three of the four train/test combinations, but did not examine frozen
encoders or individual layers. Xiao and Vu~\cite{xiao2025layerwise} trained
per-layer classifiers on frozen XLS-R with and without silence and showed,
through distributions of cosine distances, that silence affects all layers and
the shallow layers most on In-the-Wild data. They reported per-layer error
rates for only one training condition, as a figure, and did not test whether
the ordering of layers is preserved. Whether the
ranking of the layers of a frozen encoder survives silence trimming has not
been tested.

This paper examines that question with frozen WavLM Base. A two-layer
classifier is trained on mean- or statistics-pooled representations of each of
the 12 transformer layers, using ASVspoof 2019 LA audio that is either left
unchanged (\emph{clean}) or stripped of leading and trailing silence
(\emph{trimmed}), with five random seeds per configuration. Models are tested
on the ASVspoof 2019 LA evaluation set and, without any adaptation, on the
In-the-Wild dataset~\cite{mueller2022does}. For every layer, models trained on
clean audio are tested on clean audio and models trained on trimmed audio on
trimmed audio. Each model is also tested in the other condition, which yields
the complete 2$\times$2 train/test design for every layer. Layers
and checkpoints are selected on the development set only, and layer~4 is
additionally reported as a reference configuration. Uncertainty is
estimated with a speaker-level bootstrap. The aim is not a lower error rate but
an account of how much of a frozen encoder's layer-wise behavior depends on
silence. The contributions are as follows.

\begin{enumerate}
\item \textbf{Layer-wise effect of silence trimming and a test of ranking
preservation.} We report the equal error rate (EER) of every WavLM Base layer
with and without silence and test quantitatively whether the ranking of
the 12 layers survives trimming. It does so only partly: at the 40~dB
threshold, Kendall's $\tau$ between the clean and trimmed rankings is 0.58 on
the ASVspoof 2019 LA evaluation set and 0.73--0.76 on In-the-Wild, and it
decreases at the stricter thresholds of 20 and 15~dB (0.36--0.48 on the
evaluation set and 0.48--0.64 on In-the-Wild). At 40~dB the best layer does
not change (layer~5 on the
evaluation set, layer~4 on In-the-Wild), whereas layers~1 and~2 show the
largest increase in EER on the evaluation set and fall to near the bottom
of the ranking. At 15~dB, and at 20~dB with mean pooling, the best layer on the
evaluation set becomes layer~6 (with statistics pooling at 20~dB, layers 5 and 6 are
nearly equal, 5.80\% and 5.82\%), while layer~4 remains best on
In-the-Wild.

\item \textbf{Separating training-time and test-time effects of trimming.}
Earlier studies covered some or all of the four train/test combinations for
complete countermeasures~\cite{muller2021speech,wang2024silence,shim2024beyond}.
We apply the complete 2$\times$2 design to each of the 12 layers of a frozen
encoder and define each effect as an explicit difference between cells. On the
evaluation set, the EER of a model trained on clean audio rises by
6.3--19.9 percentage points when only the test audio is trimmed, with an interval excluding zero at every layer (40~dB). For layers 3--5,
the increase is 10.6--19.1 points at 20 and 15~dB. On In-the-Wild, the effect
depends on the layer. At layers 3--5, trimming the test audio lowers the
error only for models trained on trimmed audio, an interaction that is
significant in all six configurations at each of the three thresholds. At
layers 6, 8, and 9, training on trimmed audio lowers the error also on clean
test audio. At 40~dB, splitting In-the-Wild by original duration shows for
layers 3--5 that this test-time effect is concentrated in recordings no longer
than the 4-second analysis window, so the shift of the window caused by
trimming does not account for the result for these layers.

\item \textbf{Zero padding alters WavLM features.} Padding inputs to a fixed
length changes the WavLM Base representations themselves, through the group
normalization layer of its convolutional encoder. On 500 randomly sampled
training files, the mean relative deviation from unpadded features at layer~4
is 0.31 for clean and 0.44 for trimmed audio, and passing an attention mask
reduces but does not remove it. All features in this study are therefore
extracted without padding.

\item \textbf{Per-attack analysis with speaker-level uncertainty.} The effect
of trimming is reported for each of the 13 evaluation attacks, with
speaker-level bootstrap confidence intervals and alongside the out-of-domain
evaluation. At 40~dB, it ranges from an increase of 16.7--24.3 percentage points of EER for
A10 to a small but significant decrease for A17. At 20 and 15~dB, A19 also
shows a significant decrease.
\end{enumerate}

Two recommendations follow from these results. Layer-wise studies on ASVspoof
2019 LA should report both the clean and the trimmed condition, because the
ranking of the shallow layers and, at stricter trimming thresholds, the best
layer on the evaluation set depend on silence
(Sections~\ref{subsec:res-rank} and~\ref{subsec:res-threshold}). WavLM Base
features should be extracted without zero padding, so that the representation
of a file depends only on its own samples: padding changes the features
through the group normalization of the convolutional encoder, and an attention
mask does not fully undo the change. The recommendation does not rest on
performance: in the two configurations compared, padding gives a lower error
on some test sets and a higher one on others
(Section~\ref{subsec:res-padding}).

The remainder of the paper is organized as follows.
Section~\ref{sec:related} reviews related work. Section~\ref{sec:method}
describes the data, the trimming and extraction pipeline, the classifier, the
2$\times$2 design, and the statistical analysis. Section~\ref{sec:results}
presents the results, Section~\ref{sec:discussion} relates them to prior work
and states the limitations, and Section~\ref{sec:conclusion} concludes.

\section{Related Work}\label{sec:related}

\subsection{Self-supervised front ends for spoofing detection}\label{subsec:rw-ssl}

Early spoofing countermeasures combined hand-crafted cepstral features, such as
constant-Q or linear frequency cepstral coefficients, with Gaussian mixture
models or light convolutional networks
\cite{wu2015asvspoof,wang2020asvspoof,lavrentyeva2019stc}; these systems
perform well on known attacks but degrade on unseen synthesis algorithms.
End-to-end models operating on the raw waveform, notably those based on
spectro-temporal graph attention, improved substantially on this
\cite{jung2022aasist,zhang2024aasist2}. Wang and
Yamagishi~\cite{wang2022investigating} then showed that a pretrained SSL front
end outperforms hand-crafted features by a wide margin and that fine-tuning
the front end yields further gains, and subsequent systems have paired WavLM
or XLS-R with attentive, sensitive-layer, or conformer back ends
\cite{guo2024audio,zhang2024audio,rosello2023conformer}. Utterance-level
aggregation in these systems ranges from mean and statistics
pooling~\cite{snyder2018xvectors} to learned attentive
pooling~\cite{okabe2018attentive,stourbe2024exploring}; only the two fixed
variants are used here.

\subsection{Layer-wise analysis of SSL representations}\label{subsec:rw-layers}

Layer-wise analysis of a self-supervised speech model showed that acoustic
properties are most accessible in the early layers, whereas phonetic and
word-level information peaks in intermediate layers~\cite{pasad2021layerwise}.
For spoofing detection, El Kheir et al.~\cite{elkheir2025comprehensive}
analyzed several SSL models across languages and deepfake datasets and found
that lower layers carry more spoofing-relevant information than the
final ones. Pan et al.~\cite{pan2024attentive} merged the hidden embeddings of
WavLM Large with attention and reported that its lower-to-intermediate layers
contribute most to the decision; Sinha et al.~\cite{sinha2024improving}
introduced layer-conditioned embedding fusion; the Spoof-SUPERB benchmark
compared frozen SSL models with a learned weighted sum over
layers~\cite{ali2026superb}; and Beheshti et al.~\cite{beheshti2026probing}
used probing classifiers to rank layers before task-specific training,
reporting that informative layers cluster in depth zones. Other work has
characterized the temporal dynamics of these representations rather than
their layer-wise structure~\cite{zhu2023characterizing}. Two recent studies
concern frozen WavLM Base directly. Serrano et al.~\cite{serrano2025} compared
layer selection and fusion for six frozen SSL encoders on out-of-domain data
and found layer~4 to be the best single layer of WavLM Base. Mallat
et al.~\cite{mallat2026interpreting} trained a linear classifier on the
mean- and standard-deviation-pooled output of each WavLM Base layer using
ASVspoof~5, found layer~11 to perform best as a single layer, and related the layers to 17
acoustic properties through projection-weighted canonical correlation
analysis. None of these studies compares per-layer results obtained with and
without silence, although Beheshti et al.\ trim silence before training.

\subsection{Silence as a shortcut}\label{subsec:rw-silence}

M\"uller et al.~\cite{muller2021speech} showed that bonafide utterances in
ASVspoof 2019 LA carry significantly longer leading and trailing silence than
spoofed ones, that a model using only the duration of the leading silence
performs well above chance, and that trimming silence changes the error rates
of trained detectors substantially. Zhang et al.~\cite{zhang2023impact}
attributed the effect to both the proportion and the content of silence,
noted that the proportion is lower in speech generated by text-to-speech (TTS)
systems, and found that end-to-end TTS attacks remain harder to detect after
silence removal even when the detector is retrained. The ASVspoof 2021
challenge included hidden evaluation subsets from which non-speech intervals
were removed by voice activity detection, and the top-ranked systems degraded
on them, more for TTS than for voice conversion
attacks~\cite{liu2023asvspoof2021}. Shim et al.~\cite{shim2023} proposed a
general framework that introduces controlled interventions into training and
test data to expose shortcut learning in countermeasures, and Rubio
et al.~\cite{rubio2026intervention} developed an intervention-based diagnosis
for an XLS-R detector that includes manipulations of non-speech segments and
of zero padding. Methods designed to remain robust when silence is trimmed
include data augmentation with multi-task meta-learning~\cite{wang2024silence}
and multi-subspace attention~\cite{wang2026multisubspace}. These studies treat
the countermeasure as a whole and do not examine individual encoder layers.
Projecting the spoof-versus-bonafide difference in the neuron activations of
frozen XLS-R and HuBERT encoders onto a silence-versus-speech subspace, Xiao
et al.~\cite{xiao2026esp} found that this subspace explains 13--14\% of the
difference (per effective rank) on the ASVspoof 2019 evaluation set but only
0.3--0.5\% on In-the-Wild; their results are aggregated over all layers and
do not compare train/test conditions.

\subsection{Closest work}\label{subsec:rw-closest}

The studies closest to the present one are summarized in
Table~\ref{tab:closest}. Wang et al.~\cite{wang2024silence}
evaluated ten countermeasures, including fine-tuned wav2vec~2.0 systems,
under three train/test combinations: training and testing on unmodified audio,
training on unmodified and testing on trimmed audio, and training and testing
on trimmed audio, with a trimming threshold of 15~dB. They also trained a
model on silence alone and analyzed silence duration per attack, and they
tested their own method across datasets on ASVspoof 2021. Their study does not
include frozen encoders, a layer-wise analysis, or an evaluation on
In-the-Wild data. Xiao and Vu~\cite{xiao2025layerwise} attached a classifier to each
layer of a frozen XLS-R encoder, fused the per-layer decisions, and compared
training with and without silence. Using distributions of cosine distances,
they showed that silence affects all layers and the shallow layers most on
In-the-Wild data. Their training includes data augmentation, and they do not
report per-layer error rates for both training conditions, do not use a full
train/test design, and do not test whether the ordering of layers changes.
M\"uller et al.~\cite{muller2021speech} evaluated a model trained with silence
on trimmed test audio and trained RawNet2 on trimmed audio, and Shim
et al.~\cite{shim2024beyond} reported all four train/test combinations for
AASIST-L, using an energy-based detector that removes silences longer than
50~ms; none of these studies analyzes a frozen encoder layer by layer. The
present study measures the EER of every layer of a frozen encoder in all four
train/test combinations, with a fixed classifier and no augmentation, and
tests quantitatively whether the ranking of layers is preserved
(Table~\ref{tab:closest}).

\begin{table}[t]
\caption{Studies that train and test countermeasures with and without silence.
Cells are the train/test combinations of clean and trimmed audio}
\label{tab:closest}
\centering
\footnotesize
\setlength{\tabcolsep}{4pt}
\begin{tabular}{@{}lccccc@{}}
\toprule
 & M\"uller~\cite{muller2021speech} & Wang~\cite{wang2024silence}
 & Shim~\cite{shim2024beyond} & Xiao~\cite{xiao2025layerwise} & Ours \\
\midrule
Encoder            & CQT CMs, RawNet2 & 10 CMs & AASIST-L   & XLS-R       & WavLM Base \\
Frozen encoder     & ---              & ---    & ---        & \checkmark  & \checkmark \\
Per-layer EER      & ---              & ---    & ---        & figure only & \checkmark \\
Train/test cells   & 3 of 4$^{a}$     & 3 of 4 & 4 of 4     & 2 of 4      & 4 of 4 (all layers) \\
Layer-ranking test & ---              & ---    & ---        & ---         & \checkmark \\
In-the-Wild        & ---              & ---    & ---        & \checkmark  & \checkmark \\
\bottomrule
\end{tabular}
\par\smallskip
\raggedright\footnotesize $^{a}$Across different models; for RawNet2, whether
the test audio was trimmed is not stated.
\end{table}

\section{Method}\label{sec:method}

\begin{figure}[t]
\centering
\begin{tikzpicture}[
  font=\footnotesize, >={Stealth[length=2mm]},
  node distance=3.5mm and 4mm,
  box/.style={draw=fsmdeep, rounded corners=2pt, align=center,
              minimum height=10mm, text width=19mm, inner sep=2pt, text=fsmdeep},
  pre/.style={box, fill=fsmsky!18},
  opt/.style={box, fill=fsmsky!18, dashed},
  frozen/.style={box, fill=fsmslate!15, draw=fsmslate},
  train/.style={box, fill=fsmteal!12, draw=fsmteal, very thick},
  arr/.style={->, draw=fsmdeep, thick}
]
\node[frozen] (cnn) {CNN\\encoder};
\node[frozen, right=of cnn, text width=26mm] (tf)
     {12 Transformer\\layers $\mathbf{H}^{(1)},\dots,\mathbf{H}^{(12)}$};
\node[pre,   right=of tf] (sel) {Layer $l$\\$T\times768$};
\node[pre,   right=of sel] (pool) {Pooling\\mean / stats};
\node[pre, above=9mm of cnn]                  (in)   {Waveform\\16\,kHz mono};
\node[opt, above=9mm of tf, text width=26mm]  (trim) {Trim silence 40\,dB\\\emph{(trimmed)}};
\node[pre, above=9mm of sel]                  (crop) {Center crop\\to $\le$4\,s\\no padding};
\node[pre, above=9mm of pool]                 (norm) {Peak\\normalization};
\node[train, below=9mm of pool] (clf) {Classifier\\$D\to128\to2$};
\node[pre,   left=of clf] (score) {Score};
\node[draw=fsmslate, dashed, rounded corners=3pt, inner sep=2.5mm,
      fit=(cnn)(tf),
      label={[text=fsmslate,font=\footnotesize]below:WavLM Base (frozen, 94.4\,M)}] (wavlm) {};
\draw[arr] (in) -- (trim);
\draw[arr] (trim) -- (crop);
\draw[arr] (crop) -- (norm);
\draw[arr] (norm.south) |- ($(norm.south)!0.5!(norm.south|-wavlm.north)$) -| (cnn.north);
\draw[arr] (cnn) -- (tf);
\draw[arr] (tf) -- (sel);
\draw[arr] (sel) -- (pool);
\draw[arr] (pool) -- (clf);
\draw[arr] (clf) -- (score);
\node[text=fsmteal, font=\scriptsize, right=1mm of clf, align=left] {trained\\(197\,K)};
\end{tikzpicture}
\caption{Detection pipeline. Silence trimming (dashed) is applied only in the
trimmed condition. Each file enters the encoder at its own length, up to 4~s,
without zero padding. The encoder is frozen; only the classifier is trained}
\label{fig:pipeline}
\end{figure}
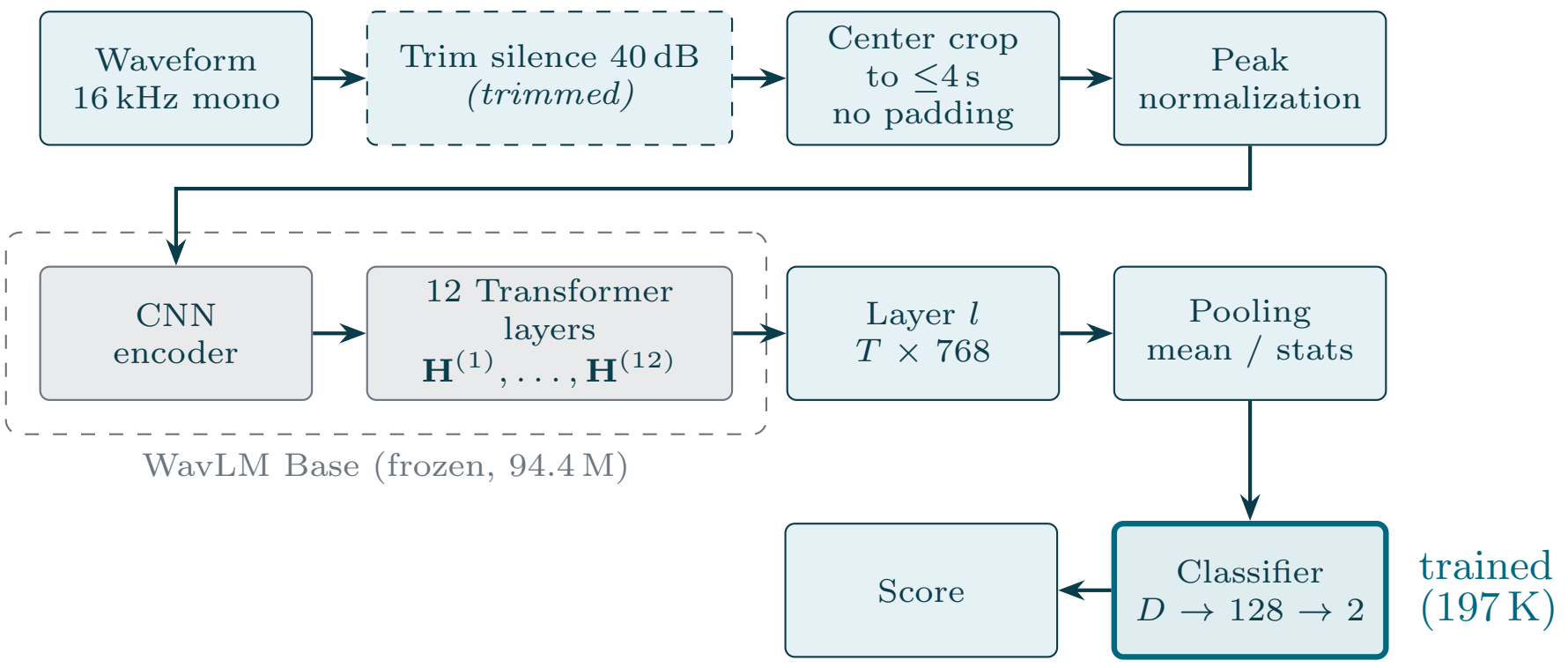

Figure~\ref{fig:pipeline} summarizes the pipeline. A waveform is optionally
trimmed of leading and trailing silence, restricted to at most 4~s without
padding, and passed through a frozen WavLM Base encoder. The hidden states of a
single layer are pooled over time and classified by a small feed-forward
network, which is the only trained component.

\subsection{Data}\label{subsec:data}

Classifiers are trained on the logical access partition of ASVspoof
2019~\cite{wang2020asvspoof} and tested on its evaluation partition and,
without any adaptation, on In-the-Wild~\cite{mueller2022does}
(Table~\ref{tab:data}). ASVspoof 2019 LA is used because the silence cue has
been documented and analyzed on this database~\cite{muller2021speech,zhang2023impact},
which allows the results to be compared directly with these studies. The
ASVspoof 2019 LA partitions are speaker-disjoint.
The training and development partitions contain six attacks (A01--A06); the
evaluation partition contains thirteen (A07--A19), of which only A16 and A19
reuse algorithms seen in training (those of A04 and A06). In-the-Wild consists
of bonafide and spoofed recordings of public figures collected from publicly
available sources, with uncontrolled channels and recording conditions; it is
never used for training or for any selection. Speaker identities, required for
the speaker-level bootstrap (Section~\ref{subsec:stats}), are taken from the
official protocol files for ASVspoof 2019 and from the dataset metadata for
In-the-Wild. The evaluation partition contains 67 speakers with bonafide files
and 48 with spoofed files; In-the-Wild contains 54 speakers in each class.

\begin{table}[t]
\caption{Datasets. Training, development and evaluation partitions are from
ASVspoof 2019 LA; In-the-Wild is used only for zero-shot testing}
\label{tab:data}
\centering
\begin{tabular}{@{}lrrrll@{}}
\toprule
Set & Bonafide & Spoof & Total & Attacks & Use \\
\midrule
Train       &  2{,}580 & 22{,}800 & 25{,}380 & A01--A06 & Classifier training \\
Development &  2{,}548 & 22{,}296 & 24{,}844 & A01--A06 & Checkpoint and layer selection \\
Evaluation  &  7{,}355 & 63{,}882 & 71{,}237 & A07--A19 & In-domain test \\
In-the-Wild & 19{,}963 & 11{,}816 & 31{,}779 & ---      & Out-of-domain test \\
\bottomrule
\end{tabular}
\end{table}

\subsection{Preprocessing and silence trimming}\label{subsec:trim}

Every file is loaded as a mono waveform at 16~kHz. Two conditions are
compared. In the \emph{clean} condition the waveform is used as is. In the
\emph{trimmed} condition, leading and trailing silence is first trimmed with
\texttt{librosa.effects.trim} (librosa~0.11.0), which discards frames at the start and end of
the file whose energy lies more than 40~dB below that of the loudest frame
(frame length 400 samples, hop 160 samples). Pauses within the utterance are
not affected. If less than 0.5~s would remain after trimming, the file is kept
untrimmed; this applied to 8 training, 5 development,
21 evaluation, and 49 In-the-Wild files. The threshold was fixed before the
experiments reported here, following M\"uller et al.~\cite{muller2021speech};
at 40~dB, the fallback to the untrimmed file applies to at most 0.15\% of the
files in any set. Both conditions then apply the same two steps. Files longer than
4~s are cropped to their central 4~s (64{,}000 samples); shorter files keep
their actual length and are not padded. Files shorter than 400 samples, the
receptive field of the convolutional encoder, are padded to 400 samples; no
such file occurs in the training, development, or In-the-Wild sets.
Finally, each waveform is peak-normalized as
$\mathbf{x}/(\max_n |x_n| + 10^{-8})$.

To examine how far the results depend on the threshold, the trimmed condition
was repeated with thresholds of 20 and 15~dB, using the same frame settings,
the same fallback rule, and the same extraction, training, and analysis.
These two thresholds were added after the 40~dB results and the measurement
of the silence remaining after 40~dB trimming were available
(Section~\ref{subsec:res-threshold}); they are therefore reported as a
sensitivity analysis, and all other results refer to 40~dB. The fallback to
the untrimmed file applied to 80, 54, 158, and 708 files at 20~dB and to 132,
105, 279, and 1{,}147 files at 15~dB (training, development, evaluation, and
In-the-Wild), that is, to at most 0.52\% of the ASVspoof 2019 LA files and
3.6\% of the In-the-Wild files. On In-the-Wild, the fallback is more frequent for bonafide than for spoofed
files (2.83\% against 1.22\% at 20~dB and 4.70\% against 1.76\% at 15~dB);
on ASVspoof 2019 LA it applies to 0.21--0.54\% of the files of either class.
Fallback files keep their silence, so at the stricter thresholds more bonafide
than spoofed In-the-Wild files retain it. No file was shorter than 400 samples at
either threshold.

\subsection{Frozen feature extraction}\label{subsec:extraction}

Features are extracted with WavLM Base~\cite{chen2022wavlm}. The publicly
released checkpoint \nolinkurl{microsoft/wavlm-base} is used; the model
consists of a convolutional encoder followed by 12 transformer layers with a
hidden dimension of 768 and about 94.4~million parameters in total; it was
pretrained on 960~hours of LibriSpeech. All encoder parameters are frozen, and
the encoder is run in evaluation mode (dropout, LayerDrop and SpecAugment
disabled) without gradient computation. For
an input of $T$ frames (one frame per 20~ms), the encoder produces hidden
states $\mathbf{H}^{(l)} \in \mathbb{R}^{T \times 768}$ for each transformer
layer $l \in \{1,\dots,12\}$; the output of the convolutional encoder is not
used. Features are extracted once per condition and cached, so every
subsequent experiment trains only the classifier.

Each file is passed through the encoder on its own (batch size~1) and without
an attention mask, so that no zero padding is ever added. This choice matters
for WavLM Base. The first layer of its convolutional encoder applies group
normalization with as many groups as channels (512), so the mean and variance
of each channel are computed over all time steps of the input, padded zeros
included. The attention mask, in contrast, acts only inside the transformer
layers and does not reach the convolutional encoder. Padding a file therefore
changes the representation of its unpadded part as well. Both standard routes
for batched input pad with zeros: the released checkpoint requests an
attention mask in its preprocessor configuration
(\texttt{return\_attention\_mask} is true) while specifying group normalization
(\texttt{feat\_extract\_norm} is \texttt{"group"})\footnote{\url{https://huggingface.co/microsoft/wavlm-base}},
and the Transformers documentation states that models with group
normalization were not trained with an attention mask and that their inputs
should simply be padded with zeros\footnote{\url{https://huggingface.co/docs/transformers/model_doc/wav2vec2}}.
The
behavior is known in software implementations of the
model\footnote{\url{https://github.com/pytorch/audio/issues/2242}}, and Rubio et al.~\cite{rubio2026intervention}
included manipulations of zero padding in their diagnosis of an XLS-R
detector. Section~\ref{subsec:res-padding} measures the effect at the
feature level, layer by layer, for an encoder with group normalization. Files that fill the 4-second window and therefore need no
padding receive exactly the same input as in padded extraction, so the
procedure differs only for shorter files. Extraction used PyTorch~2.11.0 and Hugging Face
Transformers~5.17.0 on an NVIDIA T4 GPU. The Transformers version was pinned
after it was found to change between sessions of the computing environment;
versions 5.17.0 and 5.18.0 produced identical features.

\subsection{Pooling and classifier}\label{subsec:classifier}

The frame sequence of the selected layer, $\mathbf{h}_1,\dots,\mathbf{h}_T$,
is reduced to a fixed-length vector by one of two fixed pooling operations.
Mean pooling computes
\begin{equation}
\boldsymbol{\mu} = \frac{1}{T}\sum_{t=1}^{T}\mathbf{h}_t \in \mathbb{R}^{768},
\label{eq:mean}
\end{equation}
and statistics pooling~\cite{snyder2018xvectors} appends the element-wise
standard deviation,
\begin{equation}
\boldsymbol{\sigma} = \sqrt{\frac{1}{T}\sum_{t=1}^{T}
(\mathbf{h}_t-\boldsymbol{\mu})^{2}}, \qquad
\mathbf{v}_{\mathrm{stats}} = [\boldsymbol{\mu};\boldsymbol{\sigma}]
\in \mathbb{R}^{1536}.
\label{eq:stats}
\end{equation}
The standard deviation uses the population definition (division by $T$) for
every dataset.

The pooled vector is classified by a two-layer network,
\begin{equation}
\mathrm{Linear}(D,128) \rightarrow \mathrm{ReLU} \rightarrow
\mathrm{Dropout}(0.3) \rightarrow \mathrm{Linear}(128,2),
\label{eq:clf}
\end{equation}
with $D=768$ for
mean and $D=1536$ for statistics pooling, giving 98{,}690 and 196{,}994
trainable parameters respectively. Because spoofed files outnumber bonafide
files by about nine to one in training, the cross-entropy loss is weighted by
$w_c = N/(2N_c)$, where $N$ is the number of training files and $N_c$ the
number in class $c$ (about 4.92 for bonafide and 0.56 for spoof). The
detection score is the softmax probability of the spoof class, so higher
scores indicate spoofed speech; its sign is reversed before the official
evaluation scripts, which expect higher scores for bonafide trials.

\subsection{Training and model selection}\label{subsec:training}

Each classifier is trained with Adam (learning rate $10^{-4}$, weight decay
$10^{-5}$) and a batch size of 128 for 50 epochs, with cosine annealing of the
learning rate over the 50 epochs. The development-set EER is computed after
epochs 1, 5, 10, \dots, 50 (11 evaluations), and the checkpoint with the
lowest value is retained. For this selection the EER is computed with a
ROC-based implementation that uses only the corner points of the ROC curve,
whereas every EER reported in this paper is computed with the official
ASVspoof~2019 evaluation script. Every configuration (layer, pooling, condition) is trained
with five seeds (42--46), giving $12 \times 2 \times 2 \times 5 = 240$
training runs.

Results are reported for two kinds of configuration. \emph{Dev-selected}
configurations use, for each condition and pooling method, the layer with the
lowest development-set EER averaged over seeds. \emph{Reference}
configurations are the clean condition with statistics pooling at layer~4
and the trimmed condition with mean pooling at layer~4. Layer~4 is the best single layer
reported for frozen WavLM Base on out-of-domain data by Serrano
et al.~\cite{serrano2025}, and it was the layer used in earlier versions of
this pipeline, which used zero padding. In those versions, the layer of the
clean configuration was chosen on the evaluation set, and the trimmed
configuration was the one with the lowest In-the-Wild EER. Because both
choices involved test data, the reference configurations are reported for
continuity only, and no conclusion rests on them alone; conclusions that
depend on which configuration is chosen are based on the dev-selected ones. In
the experiments reported here, no choice of epoch, layer, or configuration is
made on the evaluation set or on In-the-Wild.

\subsection{Train/test design}\label{subsec:design}

Let $E(a,b)$ denote the EER of a model trained in condition $a$ and tested in
condition $b$, where $a,b \in \{\mathrm{C},\mathrm{T}\}$ stand for clean and
trimmed. All four cells are computed for all 12 layers with both pooling
methods, which gives 24 configurations; at the stricter thresholds of
Section~\ref{subsec:res-threshold}, the cross-condition cells are computed
only for layers 3--5, which contain the reference layer (layer~4) and the
layers selected on the development set (layers 3 and 5). The checkpoint is
always selected on the development set of the training condition. The two
cross-condition cells $E(\mathrm{C},\mathrm{T})$ and $E(\mathrm{T},\mathrm{C})$
required 240 additional training runs (24 configurations, two cells, five
seeds); each of these models also scored the test data of its own training
condition, and these scores equal those of the corresponding grid model
exactly. The effects of trimming are defined as
\begin{align}
\Delta_{\mathrm{test}}  &= E(\mathrm{C},\mathrm{T}) - E(\mathrm{C},\mathrm{C}), \label{eq:dtest}\\*
\Delta_{\mathrm{train}} &= E(\mathrm{T},\mathrm{T}) - E(\mathrm{C},\mathrm{T}), \label{eq:dtrain}\\*
\Delta_{\mathrm{total}} &= E(\mathrm{T},\mathrm{T}) - E(\mathrm{C},\mathrm{C}), \label{eq:dtotal}\\*
\Delta_{\mathrm{train}}^{\mathrm{C}} &= E(\mathrm{T},\mathrm{C}) - E(\mathrm{C},\mathrm{C}), \label{eq:dtrainc}\\*
\Delta_{\mathrm{test}}^{\mathrm{T}} &= E(\mathrm{T},\mathrm{T}) - E(\mathrm{T},\mathrm{C}), \label{eq:dtestt}\\*
I &= \Delta_{\mathrm{test}}^{\mathrm{T}} - \Delta_{\mathrm{test}}
   = \Delta_{\mathrm{train}} - \Delta_{\mathrm{train}}^{\mathrm{C}}. \label{eq:inter}
\end{align}
$\Delta_{\mathrm{test}}$ is the effect of trimming only the test audio of a
model trained on clean audio; $\Delta_{\mathrm{train}}$ is the effect of also
trimming the training audio when the test audio is trimmed;
$\Delta_{\mathrm{total}} = \Delta_{\mathrm{test}} + \Delta_{\mathrm{train}}$
is the effect of trimming both; $\Delta_{\mathrm{train}}^{\mathrm{C}}$ is the
effect of trimming only the training audio, with clean test audio;
$\Delta_{\mathrm{test}}^{\mathrm{T}}$ is the effect of trimming the test audio
of a model trained on trimmed audio; and the interaction $I$ measures how much
the effect of trimming at test time depends on the training condition. All six
are expressed in percentage points of EER and are computed for the 24
configurations above.

Because the analysis window is the central 4~s of each file, trimming a file
longer than 4~s mostly shifts the window rather than removing silence from it.
To separate the two effects, In-the-Wild is additionally split by original
duration into files of at most 4~s and files longer than 4~s, and the four
cells of layers 3--5 are computed for each subset.

\subsection{Evaluation metrics}\label{subsec:metrics}

The primary metric is the EER, the operating point at which the miss rate
for bonafide trials equals the false-alarm rate for spoofed trials, computed
with the official ASVspoof~2019 script. EER is computed separately for each
seed and then averaged, and results are reported as mean $\pm$ sample
standard deviation over the five seeds.

On the ASVspoof 2019 evaluation set, the countermeasure is also assessed in
tandem with an ASV system by the minimum normalized tandem detection cost
function (t-DCF), in the formulation used for ASVspoof
2019~\cite{kinnunen2018tdcf,wang2020asvspoof}. It is computed with the official
ASVspoof~2019 script and the official ASV scores distributed with the
database. A revised formulation was introduced later~\cite{kinnunen2020tandem};
the 2019 formulation is used here so that values remain comparable with
published ASVspoof 2019 results. For a countermeasure threshold $s$,
\begin{equation}
\mathrm{t\text{-}DCF}(s) =
\frac{C_1\,P_{\mathrm{miss}}^{\mathrm{cm}}(s) + C_2\,P_{\mathrm{fa}}^{\mathrm{cm}}(s)}
     {\min(C_1, C_2)},
\label{eq:tdcf}
\end{equation}
where
$C_1 = \pi_{\mathrm{tar}}\bigl(C_{\mathrm{miss}}^{\mathrm{cm}} -
C_{\mathrm{miss}}^{\mathrm{asv}}P_{\mathrm{miss}}^{\mathrm{asv}}\bigr) -
\pi_{\mathrm{non}}C_{\mathrm{fa}}^{\mathrm{asv}}P_{\mathrm{fa}}^{\mathrm{asv}}$
and
$C_2 = C_{\mathrm{fa}}^{\mathrm{cm}}\,\pi_{\mathrm{spoof}}
\bigl(1-P_{\mathrm{miss,spoof}}^{\mathrm{asv}}\bigr)$.
The ASV error rates $P_{\mathrm{miss}}^{\mathrm{asv}}$,
$P_{\mathrm{fa}}^{\mathrm{asv}}$, and $P_{\mathrm{miss,spoof}}^{\mathrm{asv}}$
are fixed at the ASV system's EER threshold, and the minimum is taken over $s$.
The ASVspoof~2019 cost model is used: $\pi_{\mathrm{spoof}}=0.05$,
$\pi_{\mathrm{tar}}=0.9405$, $\pi_{\mathrm{non}}=0.0095$,
$C_{\mathrm{miss}}^{\mathrm{asv}}=C_{\mathrm{miss}}^{\mathrm{cm}}=1$, and
$C_{\mathrm{fa}}^{\mathrm{asv}}=C_{\mathrm{fa}}^{\mathrm{cm}}=10$. Like EER, the
minimum t-DCF is computed for each seed and reported as mean $\pm$ sample
standard deviation. In-the-Wild provides no ASV scores, so only EER is
reported for it.

\subsection{Statistical analysis}\label{subsec:stats}

The spread over seeds reflects only the randomness of training, which is small
for a frozen encoder and a simple classifier. Uncertainty due to the choice of
test files and speakers is estimated by bootstrap with 1{,}000 resamples at
two levels, both stratified by class. At the file level, files are drawn with
replacement within each class. At the speaker level, speakers are drawn with
replacement within each class and contribute all their files; this level
accounts for the similarity of files from the same speaker and gives wider,
more realistic intervals. In every resample the EER is computed for each seed
and averaged over seeds, as for the point estimates, and 95\% intervals are
taken from the 2.5th and 97.5th percentiles. Differences between two systems
or cells are computed on the same resamples (paired bootstrap) and are called
significant when their interval excludes zero.

Agreement between the layer rankings obtained in the clean-to-clean and
trimmed-to-trimmed cells is measured by Kendall's $\tau$ and Spearman's $\rho$
over the 12 layers, using the seed-averaged EER of each layer, with
speaker-level bootstrap intervals; in each resample the best layer is also
recorded. For $\tau$ and $\rho$, the percentile intervals are not centred on
the point estimate: at 40~dB, 69--84\% of the resamples give a smaller $\tau$
than the observed value (ties counted one half). Bias-corrected and
accelerated (BCa) intervals~\cite{efron1987better} computed from the same
resamples, with the acceleration estimated by leaving out one speaker at a
time within each class, are given in Table~\ref{tab:bca}. At 40~dB they lead
to the same conclusions, but their bias-correction constant ranges from
$-0.46$ to $+1.21$ over the twelve configurations of Table~\ref{tab:bca},
6--23\% of the resamples equal the observed $\tau$, which over 12 layers can
take only 67 values, and at 40~dB the upper BCa limits lie at or above the
99.8th percentile of the resamples. Percentile intervals are therefore
reported for $\tau$ and $\rho$, as for all other quantities. The exact
two-sided $p$-value of $\tau$ for $n = 12$ is also given; it treats the 12
layer EERs as fixed and tests whether their two orderings are associated,
without resampling files or speakers. Per-attack EER is computed for each evaluation attack from that
attack's spoofed files and all bonafide files of the evaluation set, per seed
and averaged over the five seeds, as everywhere else; the change caused by
trimming is tested with the same paired, class-stratified bootstrap at the
file and speaker levels.

\section{Results}\label{sec:results}

\subsection{Silence as a cue in the data}\label{subsec:res-silence}

\begin{figure}[t]
\centering
\includegraphics[width=\textwidth]{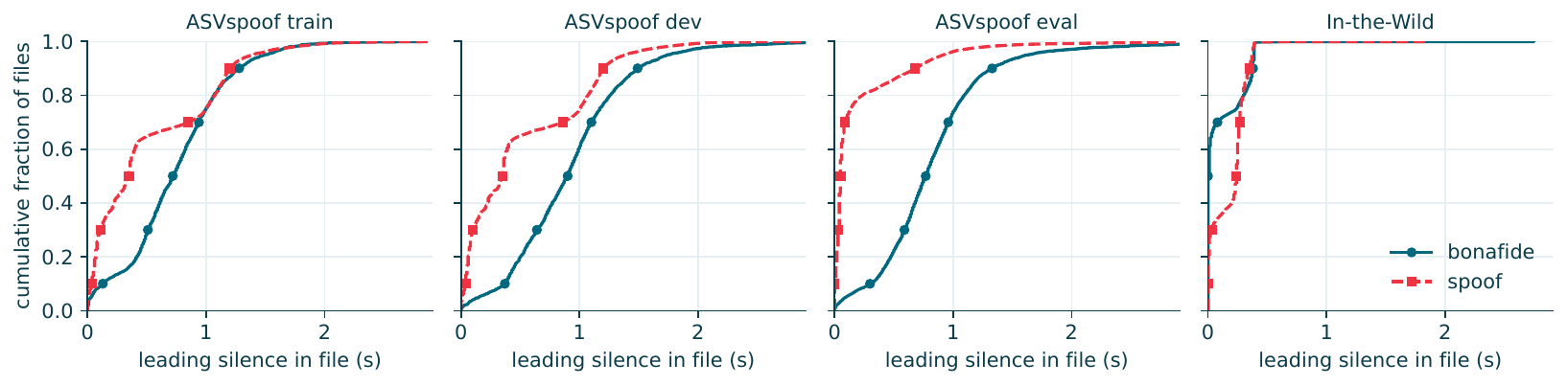}
\caption{Empirical cumulative distribution of leading silence over the whole
file (40~dB threshold) in the ASVspoof 2019 LA training, development, and
evaluation sets and in In-the-Wild, for bonafide (teal, solid, circles) and
spoofed (red, dashed, squares) files}
\label{fig:ecdf}
\end{figure}

The silence cue reported for ASVspoof 2019 LA~\cite{muller2021speech} is
present in the files used here. At the 40~dB threshold used for trimming,
the median leading silence on the evaluation set is 0.77~s for bonafide and
0.05~s for spoofed files over the whole file, and 0.72~s and 0.04~s within the
4-second analysis window; the two distributions separate clearly
(Fig.~\ref{fig:ecdf}). To quantify the cue, a logistic-regression classifier
was trained on the training partition using only silence features: leading
and trailing silence over the whole file and within the 4-second window
(\emph{sil}), optionally with the number of samples and the amount of padding
(\emph{sil+dur}). Features were log-transformed ($\log(1+x)$) and
standardized, classes were weighted inversely to their frequency, and the
audio itself was not trimmed. Without any audio model, this classifier reaches
18.82\% EER on the evaluation set when silence is measured at 30~dB and
23.40\% at 40~dB (Table~\ref{tab:silence}). On In-the-Wild, the same
classifiers give EERs between 49.20\% and 58.48\%; there, the median leading
silence is 0.00~s for bonafide and 0.24~s for spoofed files.

\begin{table}[t]
\caption{EER (\%) of a logistic-regression classifier trained on the
ASVspoof 2019 LA training partition with silence features only (sil: leading
and trailing silence over the whole file and within the 4-second window) or
with these features plus number of samples and amount of padding (sil+dur)}
\label{tab:silence}
\centering
\begin{tabular}{@{}llccc@{}}
\toprule
Features & Threshold & Development & Evaluation & In-the-Wild \\
\midrule
sil     & 30~dB & 31.90 & 18.82 & 49.20 \\
sil     & 40~dB & 30.22 & 23.40 & 58.39 \\
sil+dur & 30~dB & 31.12 & 18.75 & 49.24 \\
sil+dur & 40~dB & 34.50 & 23.25 & 58.48 \\
\bottomrule
\end{tabular}
\end{table}

\subsection{Effect of zero padding on WavLM features}\label{subsec:res-padding}

\begin{figure}[t]
\centering
\includegraphics[width=\textwidth]{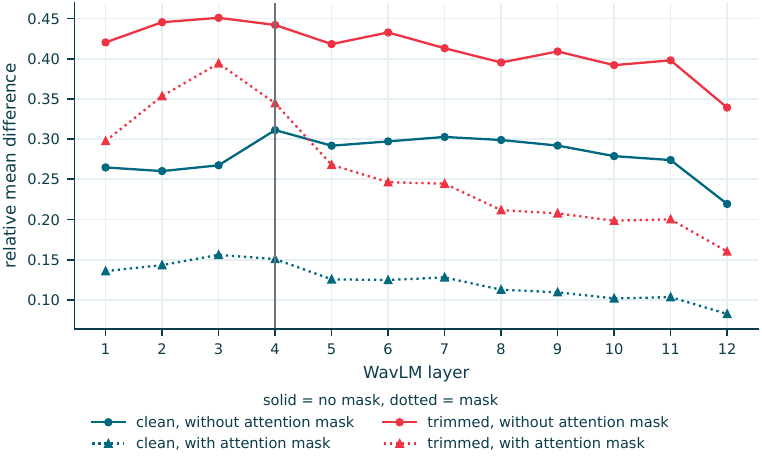}
\caption{Relative deviation of padded from unpadded WavLM Base features at
each layer, for 500 randomly sampled training files, on clean and trimmed
audio (teal: clean; red: trimmed), without (solid) and with (dotted) an
attention mask. The vertical line marks layer~4}
\label{fig:padding}
\end{figure}

To measure the effect described in Section~\ref{subsec:extraction}, features
of 500 randomly sampled training files and of the 500 most heavily padded
training files, in each condition, were extracted in three ways: without padding (reference),
padded to 4~s without an attention mask, and padded to 4~s with an attention
mask. The deviation from the reference is the mean absolute difference
divided by the mean absolute value of the reference features.
Figure~\ref{fig:padding} shows the result for every layer. On the random
files at layer~4, padding changes the features by 0.31 of their own magnitude
on clean audio and by 0.44 on trimmed audio. The attention mask reduces these
values to 0.15 and 0.34 but does not remove the effect. The effect is larger
on trimmed audio, where trimming shortens the files and increases the amount
of padding (median 1.76~s against 0.85~s on clean audio), and it grows with
the amount of padding: on trimmed audio at layer~4 without a mask, it is about
0.20 for less than 1~s of padding and about 0.94 for more than 3~s.

Padding is not a strong shortcut in itself. A classifier that uses only the
amount of padding, which depends on file duration, gives 41.6--48.9\% EER
across the training, development, evaluation, and In-the-Wild sets in both
conditions. Nor does extraction without padding lower the error. For the two
reference configurations, padded extraction (without an attention mask in the
clean condition and with one in the trimmed condition) gives an evaluation EER
of 1.23\% (clean/stats/L4) and 6.42\% (trimmed/mean/L4), against 1.48\% and
6.44\% without padding, and an In-the-Wild EER of 18.92\% and 14.00\%, against
18.26\% and 15.12\% (means over five seeds). Both pipelines use the same
classifier, training settings, seeds, checkpoint selection and official EER
script; they differ only in feature extraction. The reason for extracting without
padding is therefore not performance but that the representation of a file
then depends only on its own samples, not on the length of the analysis
window or, with batch-wise padding, on the other files in the batch.

\subsection{Layer-wise results and the 2$\times$2 design}\label{subsec:res-2x2}

\begin{figure}[t]
\centering
\includegraphics[width=\textwidth]{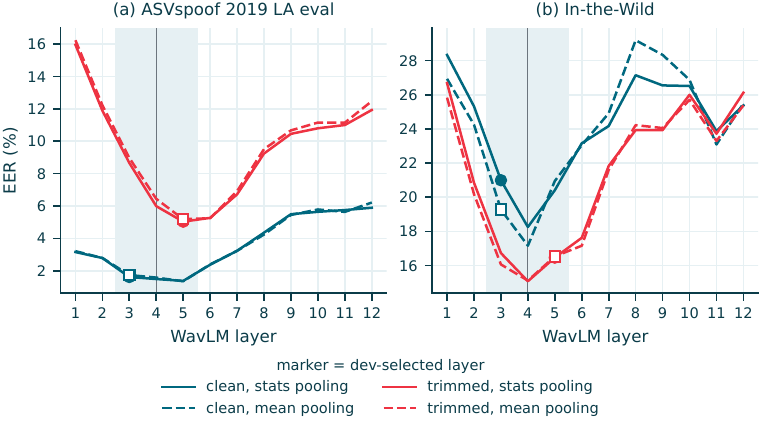}
\caption{EER of each WavLM Base layer on (a) the ASVspoof 2019 LA evaluation
set and (b) In-the-Wild, for the clean and trimmed conditions (mean over five
seeds). Teal: clean; red: trimmed. Solid: statistics pooling; dashed: mean
pooling. The shaded band marks layers 3--5 and the vertical line layer~4.
Markers indicate the layer selected on the development set (filled:
statistics; open: mean)}
\label{fig:layers}
\end{figure}

Figure~\ref{fig:layers} shows the EER of every layer in the two matched cells.
On the evaluation set, the EER of the trimmed condition exceeds that of the
clean condition at every layer, and the gap is largest in the first layers
(statistics pooling, layer~1: 3.15\% clean against 15.97\% trimmed). On
In-the-Wild the direction reverses: trimming lowers the EER at most layers,
particularly layers 2--9, and leaves layers 11 and 12 about unchanged. The
difference between mean and statistics pooling is small throughout. The three best layers are layers 3--5 in the clean condition and on
In-the-Wild in both conditions, and layers 4--6 in the trimmed condition on
the evaluation set; layer~4 has the lowest In-the-Wild EER in all four
combinations of condition and pooling.

Table~\ref{tab:main} lists the dev-selected and reference configurations.
The development set selects layer~3 in the clean condition and layer~5 in the
trimmed condition. Trimming raises the evaluation EER in all six
configurations, by 3.7--7.2 percentage points, and the increase is significant
at both the file and the speaker level in every case. The minimum tandem
t-DCF on the evaluation set changes in the same direction, from
0.0416--0.0528 in the clean condition to 0.1290--0.1667 in the trimmed
condition. On In-the-Wild,
trimming lowers the EER by 2.0--4.4 points; the decrease is significant at
the file level in all six configurations but at the speaker level only for
statistics/layer~3, mean/layer~5, and statistics/layer~5. With 54 speakers per
class, the speaker-level intervals on In-the-Wild span about $\pm$5 points.

\begin{table}[t]
\caption{Dev-selected and reference configurations: EER (\%) and minimum
tandem t-DCF as mean $\pm$ sample standard deviation over five seeds, with
95\% speaker-level bootstrap intervals for EER. Sel.: dev = dev-selected,
ref = reference configuration}
\label{tab:main}
\centering
\footnotesize
\setlength{\tabcolsep}{2pt}
\begin{tabular}{@{}llcccccc@{}}
\toprule
 & & Dev & \multicolumn{3}{c}{ASVspoof 2019 LA evaluation} &
 \multicolumn{2}{c}{In-the-Wild} \\
\cmidrule(lr){4-6}\cmidrule(l){7-8}
Configuration & Sel. & EER & EER & 95\% CI & min t-DCF & EER & 95\% CI \\
\midrule
clean/mean/L3    & dev  & 0.44 $\pm$ 0.02 & 1.73 $\pm$ 0.01 & [1.36, 2.11] & 0.0528 $\pm$ 0.0002 & 19.28 $\pm$ 0.28 & [14.34, 24.47] \\
clean/stats/L3   & dev  & 0.43 $\pm$ 0.00 & 1.61 $\pm$ 0.01 & [1.27, 1.95] & 0.0478 $\pm$ 0.0007 & 20.99 $\pm$ 0.38 & [15.60, 26.97] \\
clean/stats/L4   & ref  & 0.54 $\pm$ 0.01 & 1.48 $\pm$ 0.02 & [1.15, 1.85] & 0.0416 $\pm$ 0.0003 & 18.26 $\pm$ 0.48 & [13.64, 23.27] \\
\midrule
trimmed/mean/L4  & ref  & 1.89 $\pm$ 0.01 & 6.44 $\pm$ 0.01 & [5.47, 7.47] & 0.1667 $\pm$ 0.0007 & 15.12 $\pm$ 0.18 & [10.80, 19.99] \\
trimmed/mean/L5  & dev  & 1.60 $\pm$ 0.02 & 5.18 $\pm$ 0.02 & [4.52, 5.92] & 0.1390 $\pm$ 0.0008 & 16.54 $\pm$ 0.07 & [12.45, 21.15] \\
trimmed/stats/L5 & dev  & 1.60 $\pm$ 0.02 & 5.03 $\pm$ 0.04 & [4.29, 5.75] & 0.1290 $\pm$ 0.0006 & 16.46 $\pm$ 0.14 & [12.57, 21.06] \\
\bottomrule
\end{tabular}
\end{table}

\begin{figure}[t]
\centering
\includegraphics[width=\textwidth]{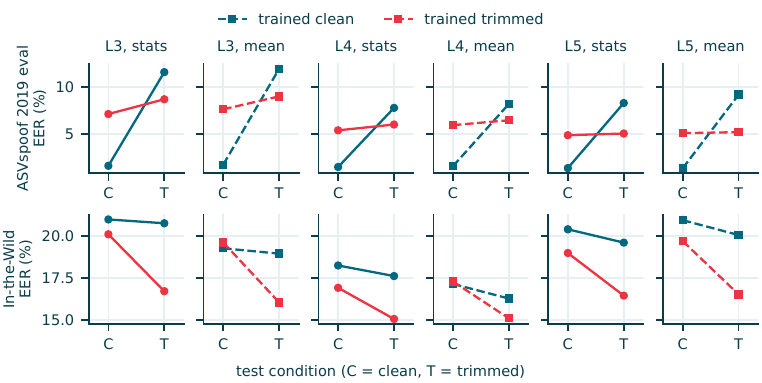}
\caption{EER in the four train/test cells for layers 3--5, which contain the
dev-selected (3 and 5) and reference (4) layers, with mean and statistics
pooling (mean over five seeds); the effects for all layers are shown in
Fig.~\ref{fig:effects-layers} and Table~\ref{tab:effects}. The horizontal axis is the test
condition; each line is one training condition (teal: trained on clean; red:
trained on trimmed). Solid: statistics pooling; dashed: mean pooling}
\label{fig:cross}
\end{figure}

\begin{figure}[t]
\centering
\includegraphics[width=\textwidth]{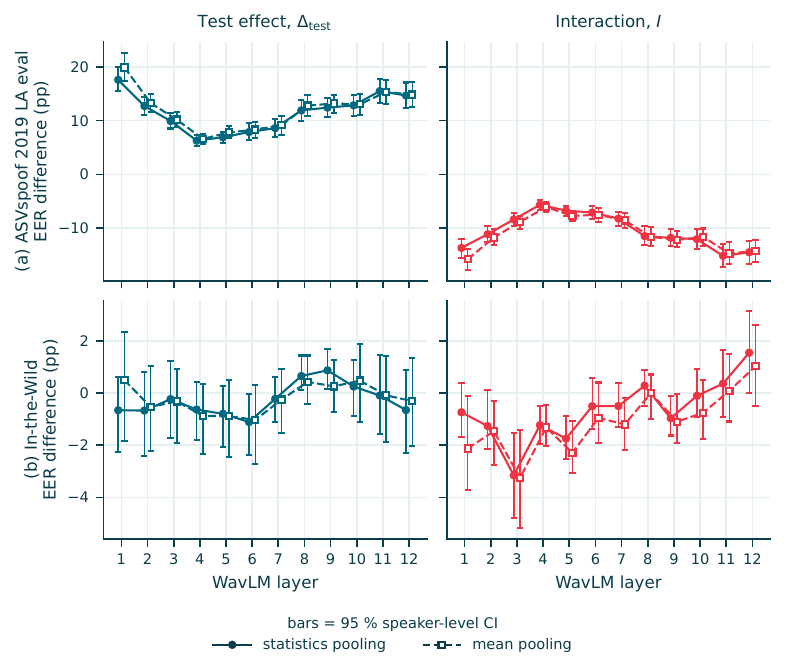}
\caption{Effects of trimming by layer on the ASVspoof 2019 LA evaluation set
(a, top row) and on In-the-Wild (b, bottom row): trimming the test audio of
the model trained on clean audio ($\Delta_{\mathrm{test}}$,
Eq.~\eqref{eq:dtest}; left column) and the interaction $I$
(Eq.~\eqref{eq:inter}; right column). Statistics pooling: solid lines, filled
circles; mean pooling: dashed lines, open squares, shifted slightly to the
right. Error bars are 95\% speaker-level paired bootstrap intervals; the
vertical axis is shared within each row. EER per seed, averaged over five
seeds}
\label{fig:effects-layers}
\end{figure}

Figure~\ref{fig:effects-layers} and Table~\ref{tab:effects} give
$\Delta_{\mathrm{test}}$ and the interaction for all 12 layers, and
Figure~\ref{fig:cross} shows the four cells for layers 3--5. On the evaluation
set, the EER of a model trained on clean audio rises by 6.27--19.88
percentage points when silence is trimmed from the test audio only
($\Delta_{\mathrm{test}}$), with an interval excluding zero in all 24
configurations; the increase is smallest at
layer~4 and largest at layer~1 with both pooling methods. A model trained on
trimmed audio changes much less when its test audio is trimmed ($+0.11$ to
$+4.13$ points; the interval excludes zero in 12 of the 24 configurations and
at neither pooling method for layers 5, 7, 9, 11, and 12;
Table~\ref{tab:effects-other}). Training on
trimmed audio while testing on clean audio raises the EER by 2.03--8.91
points, with an interval excluding zero in all 24 configurations. For
statistics/layer~4, the four
cells are $E(\mathrm{C},\mathrm{C})=1.48\%$, $E(\mathrm{C},\mathrm{T})=7.75\%$,
$E(\mathrm{T},\mathrm{C})=5.37\%$, and $E(\mathrm{T},\mathrm{T})=5.99\%$. The
interaction lies between $-5.65$ and $-15.74$ points and is negative with an
interval excluding zero in all 24 configurations. The model trained on clean
audio thus depends on silence at every layer, whereas the model trained on
trimmed audio is affected far less.

On In-the-Wild, trimming the test audio of a model trained on clean audio
changes the EER by at most 1.11 points in any of the 24 configurations; the
speaker-level interval excludes zero in three of them, with mixed sign.
Trimming both training and test audio lowers the EER with an interval
excluding zero in 11 configurations (mean pooling at layers 5--9, statistics
pooling at layers 3 and 5--9). The interaction is negative with an interval excluding zero in a different
set of 11 configurations (mean pooling at layers 1--5, 7, and 9, statistics
pooling at layers 3--5 and 9), among them all six of layers 3--5, which
contain the dev-selected (3 and 5) and reference (4) layers; in these six,
trimming the test audio lowers the EER of the model trained on trimmed audio
($\Delta_{\mathrm{test}}^{\mathrm{T}}$, by 1.85--3.56 points) but not that of the model trained on clean audio. At
layers 6, 8, and 9, training on trimmed audio lowers the EER also when the test
audio is clean ($\Delta_{\mathrm{train}}^{\mathrm{C}}$, Eq.~\eqref{eq:dtrainc}), by 2.53--5.41 points with both pooling methods, with intervals
excluding zero; at layer~7 the intervals include zero
(Table~\ref{tab:effects-other}). The
interaction is positive for statistics pooling at layer~12 ($+1.55$
[$+0.02$, $+3.14$]), and its interval includes zero in the remaining 12
configurations, among them every configuration of layers 6, 8, 10, and 11.
At layer~12, trimming the test audio of the model trained on trimmed audio
raises the EER, by 0.73 (mean) and 0.91 (statistics) points, as it does for
statistics pooling at layer~8 (0.95 points), with intervals excluding zero
(Table~\ref{tab:effects-other}).
The intervals are not adjusted for the 24 comparisons.
Table~\ref{tab:effects} reports statistics pooling; the values for mean
pooling are given in Appendix~\ref{app:mean}, together with the file- and
speaker-level intervals of layers 3--5 (Fig.~\ref{fig:forest}). For
statistics/layer~4, the cells are
$E(\mathrm{C},\mathrm{C})=18.26\%$, $E(\mathrm{C},\mathrm{T})=17.63\%$,
$E(\mathrm{T},\mathrm{C})=16.93\%$, and $E(\mathrm{T},\mathrm{T})=15.08\%$.

\begin{table}[t]
\caption{Effects of trimming (EER percentage points, statistics pooling) with
95\% speaker-level paired bootstrap intervals. Intervals that exclude zero are
in bold. Eval: ASVspoof 2019 LA evaluation set; ITW: In-the-Wild. Definitions in Eqs.~\eqref{eq:dtest}--\eqref{eq:inter}}
\label{tab:effects}
\centering
\footnotesize
\setlength{\tabcolsep}{2pt}
\begin{tabular}{@{}llcccc@{}}
\toprule
Set & L & $\Delta_{\mathrm{test}}$ & $\Delta_{\mathrm{train}}$ &
$\Delta_{\mathrm{total}}$ & $I$ \\
\midrule
Eval &  1 & \textbf{+17.58 [15.44, 19.98]} & \textbf{$-$4.76 [$-$6.67, $-$3.20]} & \textbf{+12.82 [10.97, 14.58]} & \textbf{$-$13.66 [$-$15.60, $-$12.04]} \\
     &  2 & \textbf{+12.71 [11.05, 14.39]} & \textbf{$-$3.60 [$-$5.08, $-$2.11]} & \textbf{+9.12 [7.96, 10.38]} & \textbf{$-$11.11 [$-$12.56, $-$9.64]} \\
     &  3 & \textbf{+9.94 [8.53, 11.39]} & \textbf{$-$2.88 [$-$4.09, $-$1.75]} & \textbf{+7.06 [5.98, 8.18]} & \textbf{$-$8.37 [$-$9.62, $-$7.19]} \\
     &  4 & \textbf{+6.27 [5.33, 7.28]} & \textbf{$-$1.76 [$-$2.56, $-$0.95]} & \textbf{+4.51 [3.73, 5.38]} & \textbf{$-$5.65 [$-$6.55, $-$4.78]} \\
     &  5 & \textbf{+6.90 [5.80, 7.94]} & \textbf{$-$3.24 [$-$4.10, $-$2.35]} & \textbf{+3.67 [3.06, 4.26]} & \textbf{$-$6.73 [$-$7.68, $-$5.86]} \\
     &  6 & \textbf{+7.87 [6.35, 9.53]} & \textbf{$-$4.98 [$-$6.19, $-$3.80]} & \textbf{+2.90 [2.18, 3.62]} & \textbf{$-$7.08 [$-$8.36, $-$5.87]} \\
     &  7 & \textbf{+8.60 [6.90, 10.38]} & \textbf{$-$5.11 [$-$6.41, $-$3.93]} & \textbf{+3.49 [2.67, 4.42]} & \textbf{$-$8.21 [$-$9.51, $-$6.95]} \\
     &  8 & \textbf{+11.90 [9.93, 13.82]} & \textbf{$-$7.00 [$-$8.48, $-$5.50]} & \textbf{+4.89 [3.85, 5.89]} & \textbf{$-$11.47 [$-$13.04, $-$9.67]} \\
     &  9 & \textbf{+12.41 [10.64, 14.12]} & \textbf{$-$7.44 [$-$8.74, $-$6.04]} & \textbf{+4.97 [3.89, 6.01]} & \textbf{$-$11.79 [$-$13.34, $-$10.13]} \\
     & 10 & \textbf{+12.83 [10.83, 14.75]} & \textbf{$-$7.67 [$-$9.06, $-$6.07]} & \textbf{+5.17 [3.99, 6.26]} & \textbf{$-$12.03 [$-$13.80, $-$10.21]} \\
     & 11 & \textbf{+15.47 [13.23, 17.67]} & \textbf{$-$10.22 [$-$12.08, $-$8.30]} & \textbf{+5.26 [4.05, 6.45]} & \textbf{$-$15.14 [$-$17.22, $-$12.98]} \\
     & 12 & \textbf{+14.67 [12.32, 17.08]} & \textbf{$-$8.62 [$-$10.74, $-$6.47]} & \textbf{+6.05 [4.81, 7.46]} & \textbf{$-$14.47 [$-$16.61, $-$12.39]} \\
\midrule
ITW  &  1 & $-$0.65 [$-$2.25, 0.63] & $-$0.99 [$-$6.90, 5.92] & $-$1.64 [$-$7.37, 5.08] & $-$0.73 [$-$1.68, 0.39] \\
     &  2 & $-$0.67 [$-$2.41, 0.82] & $-$3.80 [$-$8.28, 1.28] & $-$4.47 [$-$8.30, 0.05] & $-$1.26 [$-$2.13, 0.13] \\
     &  3 & $-$0.23 [$-$1.73, 1.23] & $-$4.03 [$-$7.61, 0.10] & \textbf{$-$4.26 [$-$8.01, $-$0.53]} & \textbf{$-$3.15 [$-$4.79, $-$1.52]} \\
     &  4 & $-$0.63 [$-$1.80, 0.44] & $-$2.55 [$-$5.89, 0.92] & $-$3.18 [$-$6.53, 0.29] & \textbf{$-$1.23 [$-$1.93, $-$0.48]} \\
     &  5 & $-$0.79 [$-$2.07, 0.26] & \textbf{$-$3.15 [$-$5.64, $-$0.46]} & \textbf{$-$3.94 [$-$6.54, $-$1.60]} & \textbf{$-$1.74 [$-$2.51, $-$0.87]} \\
     &  6 & \textbf{$-$1.11 [$-$2.36, $-$0.04]} & \textbf{$-$4.43 [$-$6.53, $-$1.72]} & \textbf{$-$5.53 [$-$7.28, $-$3.28]} & $-$0.50 [$-$1.36, 0.59] \\
     &  7 & $-$0.22 [$-$1.09, 0.61] & $-$2.12 [$-$4.68, 0.59] & \textbf{$-$2.34 [$-$4.71, $-$0.17]} & $-$0.49 [$-$1.29, 0.40] \\
     &  8 & \textbf{+0.66 [0.11, 1.45]} & \textbf{$-$3.87 [$-$6.71, $-$0.86]} & \textbf{$-$3.21 [$-$6.07, $-$0.15]} & +0.29 [$-$0.50, 0.89] \\
     &  9 & \textbf{+0.88 [0.15, 1.70]} & \textbf{$-$3.49 [$-$5.58, $-$1.40]} & \textbf{$-$2.61 [$-$4.80, $-$0.54]} & \textbf{$-$0.96 [$-$1.62, $-$0.12]} \\
     & 10 & +0.25 [$-$0.89, 1.27] & $-$0.77 [$-$3.48, 2.21] & $-$0.52 [$-$3.13, 2.06] & $-$0.10 [$-$0.91, 0.94] \\
     & 11 & $-$0.09 [$-$1.57, 1.46] & $-$0.06 [$-$2.97, 3.30] & $-$0.15 [$-$2.77, 2.60] & +0.37 [$-$0.92, 1.66] \\
     & 12 & $-$0.65 [$-$2.28, 0.89] & +1.39 [$-$1.72, 5.38] & +0.74 [$-$2.04, 4.08] & \textbf{+1.55 [0.02, 3.14]} \\
\bottomrule
\end{tabular}
\end{table}

\subsection{Preservation of the layer ranking}\label{subsec:res-rank}

Table~\ref{tab:rank} compares the ranking of the 12 layers in the
clean-to-clean and trimmed-to-trimmed cells. The ranking is preserved only
partly on the evaluation set ($\tau = 0.58$). On In-the-Wild, $\tau$ is
0.73--0.76, but its intervals are wide (Table~\ref{tab:rank}), so the
agreement on the two test sets is not compared directly. The best layer does not change: layer~5 on the
evaluation set and layer~4 on In-the-Wild, in both conditions and with both
pooling methods. The proportion of bootstrap resamples in which that layer is
best ranges from 55\% (trimmed, mean pooling, evaluation set) to 99\%.
At 15~dB, and at 20~dB with mean pooling, the best layer on the evaluation
set is layer~6
(Section~\ref{subsec:res-threshold}).

\begin{table}[t]
\caption{Agreement between the layer rankings in the clean-to-clean and
trimmed-to-trimmed cells, with 95\% speaker-level bootstrap intervals. The
best layer is given with the proportion of resamples in which it is best}
\label{tab:rank}
\centering
\small
\begin{tabular}{@{}llcccc@{}}
\toprule
Set & Pooling & Kendall's $\tau$ & $p$ (exact) & Spearman's $\rho$ & Best layer (clean / trimmed) \\
\midrule
Evaluation  & stats & 0.58 [0.39, 0.64] & 0.009 & 0.66 [0.52, 0.76] & L5 (82\%) / L5 (82\%) \\
Evaluation  & mean  & 0.58 [0.42, 0.64] & 0.009 & 0.66 [0.51, 0.76] & L5 (99\%) / L5 (55\%) \\
In-the-Wild & stats & 0.73 [0.42, 0.85] & 0.0005 & 0.90 [0.57, 0.94] & L4 (90\%) / L4 (82\%) \\
In-the-Wild & mean  & 0.76 [0.39, 0.82] & 0.0002 & 0.87 [0.51, 0.94] & L4 (91\%) / L4 (69\%) \\
\bottomrule
\end{tabular}
\end{table}

The disagreement comes from the first layers (Table~\ref{tab:rankdetail}).
On the evaluation set with statistics pooling, layers 1 and 2 show the largest
increase in EER after trimming (12.8 and 9.1 points) and fall from ranks 6
and 5 to ranks 12 and 10. Layers 7--12 roughly double their EER
($\times$1.9--2.1) and keep their relative order.

\begin{table}[t]
\caption{EER (\%) per layer on the evaluation set, statistics pooling, in the
clean-to-clean and trimmed-to-trimmed cells (mean over five seeds)}
\label{tab:rankdetail}
\centering
\small
\begin{tabular}{@{}lcccc@{}}
\toprule
Layer & Clean & Trimmed & Ratio & Rank (clean $\rightarrow$ trimmed) \\
\midrule
1      & 3.15 & 15.97 & $\times$5.1 & 6 $\rightarrow$ 12 \\
2      & 2.78 & 11.90 & $\times$4.3 & 5 $\rightarrow$ 10 \\
3      & 1.61 & 8.67  & $\times$5.4 & 3 $\rightarrow$ 5 \\
4      & 1.48 & 5.99  & $\times$4.0 & 2 $\rightarrow$ 3 \\
5      & 1.37 & 5.03  & $\times$3.7 & 1 $\rightarrow$ 1 \\
6      & 2.37 & 5.26  & $\times$2.2 & 4 $\rightarrow$ 2 \\
7--12  & 3.23--5.89 & 6.72--11.94 & $\times$1.9--2.1 & same relative order \\
\bottomrule
\end{tabular}
\end{table}

\subsection{In-the-Wild by original duration}\label{subsec:res-duration}

Table~\ref{tab:duration} splits In-the-Wild into files of at most 4~s
(19{,}387) and files longer than 4~s (12{,}392). For files longer than the
window, trimming mostly shifts the window, and trimming only the test audio
has no effect ($\Delta_{\mathrm{test}} = -0.08 \pm 0.09$). For files that fit
in the window, it lowers the EER by $0.90 \pm 0.08$ points (mean and
standard deviation over the five seeds), but the speaker-level interval of
this effect includes zero, here and at layers 3 and 5 with both pooling
methods. For models trained on trimmed audio, trimming the test audio lowers
the EER of the short files with a speaker-level interval that excludes zero
in all of these configurations. The test-time effect is therefore
concentrated in the short files, clearly so for models trained on trimmed
audio, and the shift of the window does not account for the results. The two subsets also differ in what is
needed for an improvement. For long files, training on trimmed audio is
enough: $E(\mathrm{T},\mathrm{C}) = 12.81\%$ against
$E(\mathrm{C},\mathrm{C}) = 16.65\%$. For short files, the improvement requires trimming in both training and test. The same pattern holds
for layers 3 and 5 and for mean pooling.

\begin{table}[t]
\caption{In-the-Wild split by original duration, statistics pooling, layer~4.
EER (\%) per cell and effects in percentage points (mean $\pm$ sample standard
deviation over five seeds)}
\label{tab:duration}
\centering
\small
\begin{tabular}{@{}lcc@{}}
\toprule
 & $\le$4~s (19{,}387 files) & $>$4~s (12{,}392 files) \\
\midrule
$E(\mathrm{C},\mathrm{C})$ & 20.34 & 16.65 \\
$E(\mathrm{C},\mathrm{T})$ & 19.44 & 16.57 \\
$E(\mathrm{T},\mathrm{C})$ & 19.65 & 12.81 \\
$E(\mathrm{T},\mathrm{T})$ & 17.43 & 12.70 \\
\midrule
$\Delta_{\mathrm{test}}$  & $-$0.90 $\pm$ 0.08 & $-$0.08 $\pm$ 0.09 \\
$\Delta_{\mathrm{train}}$ & $-$2.01 $\pm$ 0.30 & $-$3.88 $\pm$ 0.26 \\
$\Delta_{\mathrm{total}}$ & $-$2.91 $\pm$ 0.26 & $-$3.95 $\pm$ 0.27 \\
\bottomrule
\end{tabular}
\end{table}

\subsection{Per-attack analysis}\label{subsec:res-attack}

\begin{figure}[t]
\centering
\includegraphics[width=\textwidth]{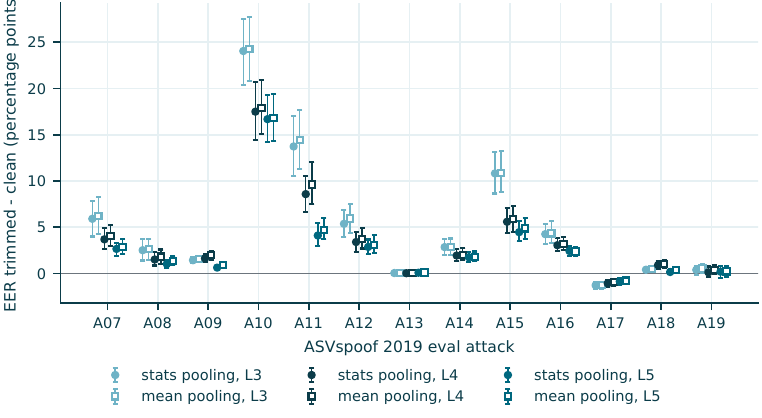}
\caption{Change in EER after trimming (trimmed minus clean) for each
evaluation attack and the six configurations, with 95\% speaker-level
bootstrap intervals. Colors indicate the layer (sky teal: L3; deep teal: L4;
teal: L5); filled circles: statistics pooling; open squares: mean pooling}
\label{fig:attack}
\end{figure}

Trimming does not affect all attacks equally (Fig.~\ref{fig:attack},
Table~\ref{tab:attack}). Each attack contributes 4{,}914 spoofed files from 48
speakers, scored against all bonafide files of the evaluation set. A10 is the
most affected, with an increase of 16.7--24.3 points across the six
configurations, from about 3--4.5\% to 20--29\% EER. A11 and A15 follow, with
increases of 4.1--14.4 and 4.5--10.9 points. A07, A08, A09, A12, A14, and A16
increase by 0.6--6.3 points, significantly in all six configurations. The
remaining changes are small: A13 by $+0.04$ to $+0.12$ points (significant
except for statistics/layer~4), A18 by $+0.2$ to $+1.0$ (not significant for
statistics/layer~5), and A19 by $+0.1$ to $+0.5$ (significant only for
mean/layer~3). A17 is the only attack whose EER decreases after trimming, by
0.76--1.27 points, significantly at both levels in all six configurations.
For most attacks the increase becomes smaller with depth from layer~3 to
layer~5; with statistics pooling, for example, it is 13.7, 8.6, and 4.1 points
for A11 and 10.8, 5.6, and 4.5 points for A15. The exceptions are A09 and
A18, whose largest increase is at layer~4, and A13 and A19, whose changes are
very small. These attacks are therefore the ones whose separation from
bonafide speech depends most on silence in the clean condition.

\begin{table}[t]
\caption{Change in EER after trimming (trimmed minus clean, percentage points)
by attack; ranges over the six configurations}
\label{tab:attack}
\centering
\small
\begin{tabular}{@{}lll@{}}
\toprule
Group & Attacks & Change \\
\midrule
Most affected  & A10 & +16.7 to +24.3 \\
Large          & A11, A15 & +4.1 to +14.4; +4.5 to +10.9 \\
Moderate       & A07, A08, A09, A12, A14, A16 & +0.6 to +6.3 \\
Small          & A13; A18; A19 & +0.04 to +0.12; +0.2 to +1.0; +0.1 to +0.5 \\
Decrease       & A17 & $-$0.76 to $-$1.27 \\
\bottomrule
\end{tabular}
\end{table}

\subsection{Sensitivity to the trimming threshold}\label{subsec:res-threshold}

Trimming at 40~dB weakens the silence cue but does not remove it. Measured on
the trimmed audio, the median leading silence that remains on the evaluation
set is 0.16~s for bonafide and 0.03~s for spoofed files when silence is
defined by a 15~dB threshold, and 0.14~s and 0.02~s at 20~dB. The
silence-only classifier of Section~\ref{subsec:res-silence}, trained on these
residual measurements, reaches 32.67\%, 33.48\%, and 35.58\% EER on the
evaluation set with silence measured at 15, 20, and 30~dB (sil), and
31.55\%--34.79\% with sil+dur, compared with 18.82\% and 23.40\% on the
untrimmed audio (Table~\ref{tab:silence}). On In-the-Wild, the same
classifiers give EERs between 33.40\% and 45.62\%. After trimming at 40~dB,
more than 50~ms of leading silence remains in 3.1--4.2\% of the files of each
set (2.1--4.0\% within the 4-second window), more often in bonafide than in
spoofed files (6.7\% and 3.0\% on the evaluation set), with a median of
0.16--0.23~s. This remainder arises when a brief sound just above the threshold
precedes a pause: the cut is placed at that sound, and the pause after it is
kept. Five bonafide In-the-Wild files are too quiet to be trimmed at 40~dB
(loudest-frame RMS below $10^{-3}$, so that librosa's floor of $10^{-5}$ lies
less than 40~dB below it).

The stricter thresholds remove more audio from both classes. The median
duration removed from an evaluation file grows from 1.17~s at 40~dB to
1.56~s at 20~dB and 1.64~s at 15~dB for bonafide files, and from 0.29~s to
0.45~s and 0.51~s for spoofed files; at 15~dB, this is a median of 54\% of the
duration of a bonafide file. On In-the-Wild, where the 40~dB rule removes a
median of 0.01~s from bonafide and 0.35~s from spoofed files, the stricter
thresholds remove 0.71~s and 0.63~s at 20~dB and 0.78~s and 0.71~s at 15~dB.
The additional audio removed at the stricter thresholds begins where the
40~dB rule detects sound and is therefore not necessarily silence.

Table~\ref{tab:threshold} repeats the main analyses with thresholds of 20 and
15~dB. In the trimmed condition, the dev-selected layer is layer~5 for both
pooling methods at all three thresholds; in the clean condition, it is
layer~3 (Table~\ref{tab:main}). With statistics pooling, the EER after trimming is
higher than on clean audio at every layer and every threshold, and the EER of
the dev-selected configuration on the evaluation set rises from
5.03\% at 40~dB to 5.80\% at 20~dB and 6.20\% at 15~dB. At the stricter
thresholds, the cross-condition cells were computed only for layers 3--5,
which contain the dev-selected and reference layers. For these layers, the
test-time effect on the evaluation set grows from 6.3--10.2 points at 40~dB
to 10.6--18.0 points at 20~dB and 11.2--19.1 points at 15~dB, and the
interaction for
statistics pooling at layer~4 from $-5.65$ to $-10.80$ and $-11.15$ points.
The layer rankings agree less at the stricter thresholds (Kendall's $\tau$
with exact $p$-values in Table~\ref{tab:bca}; at 20 and 15~dB, the exact $p$
exceeds 0.05 for three of the four evaluation-set configurations, so that for
these three the exact test does not distinguish the agreement from that of two
unrelated orderings, whereas the percentile intervals exclude zero), and the
best layer
on the evaluation set changes from layer~5 to layer~6 at 15~dB with both
pooling methods (in 63\% and 89\% of the resamples) and at 20~dB with mean
pooling (95\%); at 20~dB with statistics pooling, layers 5 and 6 are
nearly equal (5.80\% and 5.82\%), and layer~6 is best in 56\% of the resamples. On In-the-Wild, the best layer
remains layer~4, in 98\%--99\% of the resamples.

On In-the-Wild, the interaction remains negative, with speaker-level
intervals that exclude zero in all six configurations of layers 3--5 at both
stricter thresholds. Trimming the test audio alone, however, now raises the
EER of the model trained on clean audio, by 0.3--2.5 points at 20~dB and
1.0--3.7 points at 15~dB, whereas at 40~dB it lowered it by 0.2--0.9 points
for these layers.
The speaker-level intervals of this effect include zero except for mean
pooling at layer~5 (20 and 15~dB) and at layer~3 (15~dB).

The ordering of the attacks is largely unchanged. A10 remains the attack most
affected by trimming, with an increase of 19.7--32.2 points at 20~dB and
20.5--32.2 points at 15~dB. A17 is detected better after trimming at all
three thresholds. A19, whose EER rises slightly after trimming at 40~dB, is
detected better after trimming at 20 and 15~dB in all six configurations,
with speaker-level intervals that exclude zero.

\begin{table}[t]
\caption{Main results at the three trimming thresholds. Effects are in
percentage points with 95\% speaker-level bootstrap intervals (statistics
pooling, layer~4); Kendall's $\tau$ compares the clean and trimmed layer
rankings. Fallback files are those kept untrimmed because less than 0.5~s
would remain. Median removed is the median duration removed from a file by
trimming, with fallback files counted as zero. The best layer is given with the proportion of resamples in
which it is best (statistics pooling). Cross-condition cells, and hence
$\Delta_{\mathrm{test}}$ and $I$, are available for layers 3--5 only at 20
and 15~dB}
\label{tab:threshold}
\centering
\footnotesize
\setlength{\tabcolsep}{4pt}
\begin{tabular}{@{}lccc@{}}
\toprule
 & 40~dB & 20~dB & 15~dB \\
\midrule
Fallback files, eval / ITW & 21 / 49 & 158 / 708 & 279 / 1{,}147 \\
Median removed (s), eval, bona / spoof & 1.17 / 0.29 & 1.56 / 0.45 & 1.64 / 0.51 \\
Median removed (s), ITW, bona / spoof  & 0.01 / 0.35 & 0.71 / 0.63 & 0.78 / 0.71 \\
Dev-selected layer, trimmed & L5 & L5 & L5 \\
Eval EER, stats, L5 (\%) & 5.03 & 5.80 & 6.20 \\
ITW EER, stats, L5 (\%) & 16.46 & 16.42 & 16.67 \\
\midrule
Kendall's $\tau$, eval, stats / mean & 0.58 / 0.58 & 0.48 / 0.39 & 0.42 / 0.36 \\
Kendall's $\tau$, ITW, stats / mean & 0.73 / 0.76 & 0.52 / 0.58 & 0.64 / 0.48 \\
Best layer, eval & L5 (82\%) & L5 (44\%) & L6 (63\%) \\
Best layer, ITW & L4 (82\%) & L4 (99\%) & L4 (98\%) \\
\midrule
$\Delta_{\mathrm{test}}$, eval, L3--L5, both poolings & 6.3 to 10.2 & 10.6 to 18.0 & 11.2 to 19.1 \\
$\Delta_{\mathrm{total}}$, eval & 4.51 [3.73, 5.38] & 6.59 [5.57, 7.69] & 6.79 [5.68, 7.98] \\
Interaction, eval & $-$5.65 [$-$6.55, $-$4.78] & $-$10.80 [$-$12.35, $-$9.32] & $-$11.15 [$-$12.70, $-$9.66] \\
$\Delta_{\mathrm{total}}$, ITW & $-$3.18 [$-$6.53, 0.29] & $-$4.55 [$-$8.15, $-$0.80] & $-$4.06 [$-$7.93, 0.002] \\
Interaction, ITW & $-$1.23 [$-$1.93, $-$0.48] & $-$3.83 [$-$5.24, $-$2.23] & $-$3.47 [$-$4.95, $-$1.80] \\
\bottomrule
\end{tabular}
\end{table}

\subsection{What the layers encode}\label{subsec:res-probe}

\begin{figure}[t]
\centering
\includegraphics[width=\textwidth]{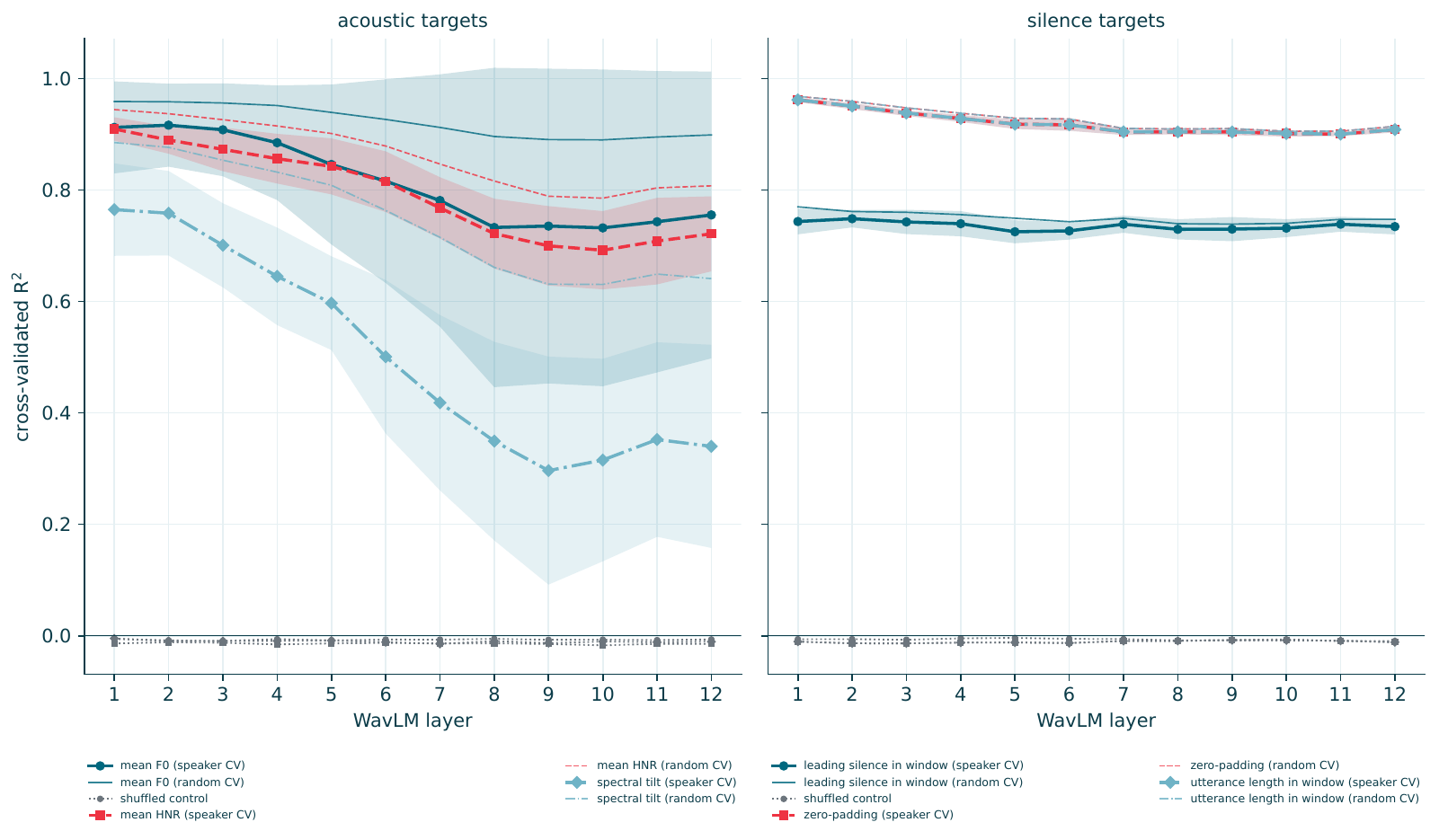}
\caption{Probe $R^2$ per layer on clean audio. Left: acoustic targets; right:
silence and length targets. Lines show the mean cross-validated $R^2$ of ridge
probes on mean-pooled features; thick lines with markers: speaker-grouped
5-fold CV, with shaded bands of $\pm$1 SD across the five folds; thin lines:
random 5-fold CV; grey dotted lines: shuffled-target control. Because
padding equals the 4~s window minus the length, the padding and length
targets in the right panel have identical $R^2$}
\label{fig:probes}
\end{figure}

Linear probes were trained to predict acoustic and silence properties from the
mean-pooled features (768 dimensions) of each layer, using 5{,}000 training files from 20 speakers,
stratified by label and attack. The acoustic targets, mean F0, mean
harmonics-to-noise ratio (HNR), and spectral tilt, were computed with Praat
through the Parselmouth library. Ridge regression was evaluated with
five-fold cross-validation grouped by speaker, and probes trained on shuffled targets
gave $R^2$ close to zero. Figure~\ref{fig:probes} and Table~\ref{tab:probes}
summarize the results on clean audio. Acoustic properties are best predicted
from layer~1 and less well with increasing depth, consistent with earlier
layer-wise analysis of SSL models~\cite{pasad2021layerwise}. Leading silence
is predicted equally well from every layer ($R^2 \approx 0.74$), so every
layer carries the silence cue. The length of the signal within the 4-second
window is also predicted well at every layer, with $R^2$ between 0.90 and
0.96 on clean audio and at least 0.88 on trimmed audio, yet a classifier based on the amount of padding, which is the
window length minus this length, is close to chance
(Section~\ref{subsec:res-padding}). The probes are descriptive: the layers
that encode acoustic properties best (layer~1) are not the layers that detect
spoofing best (layers 3--5), so they do not explain the layer ranking. Mallat
et al.~\cite{mallat2026interpreting}, who analyzed frozen WavLM Base trained
on ASVspoof~5 with a different analysis method, found most acoustic
properties to peak around layer~7.

\begin{table}[t]
\caption{Probe $R^2$ at selected layers (clean audio, speaker-grouped
cross-validation)}
\label{tab:probes}
\centering
\small
\begin{tabular}{@{}lcccc@{}}
\toprule
Target & L1 & L4 & L8 & L12 \\
\midrule
Mean F0            & 0.91 & 0.88 & 0.73 & 0.76 \\
Mean HNR           & 0.91 & 0.86 & 0.72 & 0.72 \\
Spectral tilt      & 0.76 & 0.65 & 0.35 & 0.34 \\
Leading silence    & 0.74 & 0.74 & 0.73 & 0.73 \\
Length in the 4~s window & 0.96 & 0.93 & 0.90 & 0.91 \\
\bottomrule
\end{tabular}
\end{table}

\section{Discussion}\label{sec:discussion}

\subsection{Relation to prior work on silence}\label{subsec:disc-silence}

The size of the shortcut agrees with earlier reports. M\"uller
et al.~\cite{muller2021speech} obtained 15.12\% EER on the ASVspoof 2019
evaluation set from the duration of the leading silence alone, and the EER of
RawNet2 rose from 3.61\% to 15.50\%, about fourfold, when silence was
trimmed for training; their description does not make clear whether the
evaluation audio was also trimmed. Here, silence features alone give 18.82\%
EER at 30~dB (Table~\ref{tab:silence}). With statistics pooling, training on
trimmed audio raises the EER of layers 3--5 by a factor of 3.7--5.4 when the
test audio is also trimmed (Table~\ref{tab:rankdetail}) and by a factor of
3.6--4.4 when it is clean (Tables~\ref{tab:rankdetail}
and~\ref{tab:effects-other}), so the comparison holds under either reading. The
attacks most affected by trimming also match. Zhang
et al.~\cite{zhang2023impact} likewise found the largest degradation for A10
and A11 and attributed it to their Tacotron~2 acoustic model, whose outputs
lack leading and trailing silence. The ASVspoof 2021 hidden subsets showed a
larger degradation for TTS than for voice conversion
attacks~\cite{liu2023asvspoof2021}, the latter being A17--A19 in ASVspoof 2019
LA. In this study, A10, A11, and A15 are the most
affected, A18 and A19 change little at 40~dB, and A17 improves
(Section~\ref{subsec:res-attack}); at 20 and 15~dB, A19 also improves
(Section~\ref{subsec:res-threshold}).

On In-the-Wild the cue runs the other way: at 40~dB, the median leading
silence is 0.00~s for bonafide and 0.24~s for spoofed files
(Section~\ref{subsec:res-silence}, Fig.~\ref{fig:ecdf}), and the silence-only
classifier trained on ASVspoof 2019 LA gives EERs between 49.20\% and
58.48\% there (Table~\ref{tab:silence}), close to or worse than chance.

The 2$\times$2 results extend the three-cell comparison of Wang
et al.~\cite{wang2024silence} and the four-cell comparison of Shim
et al.~\cite{shim2024beyond} for AASIST-L to every layer of a frozen encoder.
The cell of training on trimmed and testing on clean audio separates the
effect of the training condition from that of the test condition. On the evaluation set it shows that a model
trained without silence gains little from silence at test time
($\Delta_{\mathrm{test}}^{\mathrm{T}}$ is $+0.11$ to $+4.13$ points), whereas the model trained with silence depends
on it at every layer. On In-the-Wild, the result depends on the layer: at
layers 3--5, neither training nor testing on trimmed audio alone changes the
EER significantly at the speaker level, while the interaction is significant
in all six configurations; at layers 6, 8, and 9, training on trimmed audio
alone lowers the EER significantly. The layer-wise observation of
Xiao and Vu~\cite{xiao2025layerwise}, that silence affects the shallow layers
of frozen XLS-R most, is consistent with the behavior of layers 1 and 2 of
WavLM Base here, which show the largest increase in EER and the largest drop
in rank after trimming.

\subsection{Layer selection}\label{subsec:disc-layers}

The best layers lie between layers 3 and 6 in every condition, and layer~4
has the lowest In-the-Wild EER in all four combinations of condition and
pooling.
This agrees with Serrano et al.~\cite{serrano2025}, who found layer~4 to be
the best single layer of frozen WavLM Base on out-of-domain data with a
different training corpus (ASVspoof~5), input length (3-second segments), and
pooling (mean). Mallat et al.~\cite{mallat2026interpreting}, who also used
frozen WavLM Base but trained on ASVspoof~5 with a linear classifier per
layer, found layer~11 to perform best as a single layer; the two studies differ in training data
and classifier, and the present results do not identify which difference is
responsible. The probing results show that acoustic properties are most
accessible in layer~1, in line with Pasad et al.~\cite{pasad2021layerwise},
so they do not account for the advantage of layers 3--5.

Because the best layers lie in the lower half of the encoder, the layers above
them need not be computed at inference time. El Kheir
et al.~\cite{elkheir2025comprehensive} showed that keeping only the lower
layers of SSL front ends (4--6 for base and 10--12 for large models) gives
EERs comparable to the full model while at least halving the parameters.

\subsection{Out-of-domain performance in context}\label{subsec:disc-itw}

The In-the-Wild EERs of the dev-selected configurations lie
between 16.46\% and 20.99\%. For frozen encoders, Serrano
et al.~\cite{serrano2025} reported 17.8\% for layer~4 of WavLM Base trained on
ASVspoof~5, and the Spoof-SUPERB benchmark~\cite{ali2026superb} reported
24.33\% for WavLM Large and 27.28\% for HuBERT Base, both trained on ASVspoof
2019 LA and each from a single run with a fixed seed, whereas the values here
are means over five seeds. Considerably lower error rates have been reported with larger encoders and
different training: Beheshti et al.~\cite{beheshti2026probing} reached 4.94\%
with four selected layers of XLS-R-300M, data augmentation, and silence
trimming before training (9.68\% with WavLM Large in the same study, from a single run), and
Pascu et al.~\cite{pascu2024towards} reached 7.2\% with frozen
wav2vec2/xls-r-2b representations and logistic regression. The aim of the
present study is not a lower error rate but an account of how the silence
shortcut affects each layer of a frozen encoder, and the configurations were
chosen to make that account interpretable rather than to maximize
out-of-domain performance. In particular, the encoder was kept frozen and the
classifier small so that differences between configurations can be attributed
to the layer and the training condition.

\subsection{Implications}\label{subsec:disc-implications}

Two practical points follow. First, at the 40~dB threshold, layer-wise
conclusions drawn on ASVspoof 2019 LA with silence present are stable for the
best layer and the deepest layers but not for the shallowest ones, whose rank
depends strongly on silence; at stricter thresholds, the best layer on the
evaluation set also changes, to the adjacent layer, in three of the four
configurations. Reporting both conditions
makes this visible. Second, any pipeline
that pads WavLM Base inputs to a fixed length alters the features through the
group normalization of the convolutional encoder, and passing an attention
mask does not prevent it; extracting each file at its own length avoids the
problem at the cost of batching. The effect was measured here for WavLM Base
only, whose convolutional encoder uses group normalization; whether other
encoders are affected depends on the normalization they use and was not
tested.

\subsection{Limitations}\label{subsec:limitations}

The study has several limitations. Only one encoder, WavLM Base, was
examined, and the results may differ for larger models or other SSL encoders.
In-the-Wild contains 54 speakers per class, so the speaker-level intervals on
it are wide, about $\pm$5 percentage points. The main analysis uses a single
trimming threshold (40~dB); the stricter thresholds of 20 and 15~dB, the
latter equal to the threshold used by Wang et al.~\cite{wang2024silence},
were added after the 40~dB results were known and are reported as a
sensitivity analysis. Trimming at 40~dB leaves part of the silence cue in
place (Section~\ref{subsec:res-threshold}), and the silence that remains after
trimming at 20 and 15~dB was not measured; the additional audio removed at
20 and 15~dB begins where the 40~dB rule detects sound
(Section~\ref{subsec:res-threshold}) and may include weak speech sounds or
breaths at the start and end of an utterance, which were not inspected, so at
these thresholds the trimmed condition may remove more than silence; at these
thresholds the cross-condition cells were computed only for layers 3--5. A single analysis
window (4~s) was used. Only leading and
trailing silence was removed, which is the form of the cue documented for this
database~\cite{muller2021speech}; pauses within utterances were not
examined. The classifier is a small network on a frozen
encoder, and fine-tuned systems may behave differently. All models were
trained on ASVspoof 2019 LA only.

\section{Conclusion}\label{sec:conclusion}

This paper examined how the silence shortcut in ASVspoof 2019 LA affects the
layer-wise behavior of a frozen WavLM Base encoder. Trimming leading and
trailing silence raises the in-domain EER at every layer and preserves the
ranking of the 12 layers only partly. The best layer is unchanged at 40~dB but
changes on the evaluation set at stricter thresholds in three of the four
configurations, while the layer
selected on the development set in the trimmed condition is layer~5 at all
three (layer~3 in the clean condition). The complete 2$\times$2
design shows that a model trained with silence depends on it in domain at
every layer, and that on In-the-Wild the effect of trimming depends on the
layer. Zero padding alters WavLM Base features through group normalization,
so all features were extracted without padding.

Two recommendations follow. Layer-wise studies on ASVspoof 2019 LA should
report both the clean and the trimmed condition, because the ranking of the
shallow layers and, at stricter trimming thresholds, the best layer on the
evaluation set depend on silence. WavLM Base features should be extracted
without zero padding, so that the representation of a file depends only on
its own samples: padding changes the features through the group normalization
of the convolutional encoder, and an attention mask does not fully undo the change.
The recommendation does not rest on performance: in the two configurations
compared, padding gives a lower error on some test sets and a higher one on
others. Repeating the analysis with other SSL encoders and training
corpora, and measuring the silence that remains after stricter trimming,
would show how far these findings generalize.

\backmatter

\section*{Declarations}

\textbf{Funding.} This research did not receive any specific grant from funding
agencies in the public, commercial, or not-for-profit sectors.

\textbf{Competing interests.} The authors declare that they have no competing
interests.

\textbf{Ethics approval and consent to participate.} Not applicable. This
study uses only the publicly available ASVspoof 2019 Logical Access and
In-the-Wild datasets; no new data were collected from human participants.

\textbf{Consent for publication.} Not applicable.

\textbf{Materials availability.} Not applicable.

\textbf{Data availability.} The ASVspoof 2019 Logical Access database is
publicly available at \url{https://datashare.ed.ac.uk/handle/10283/3336}. The
In-the-Wild dataset is available at
\url{https://deepfake-total.com/in_the_wild}.

\textbf{Code availability.} The code, the configuration files and the result
tables behind the figures, tables and reported numbers of this article are
available at \url{https://github.com/Nooralabd/silence-shortcut-wavlm-layers}
(release v1.0) under the MIT License.

\textbf{Use of generative AI.} During the preparation of this work, the
authors used Claude Opus 5.5 (Anthropic) in order to improve the language and
readability of the manuscript, to organize and document the code, and to
support the literature search and the checking of the manuscript for internal
consistency. The study design, the experiments, the analyses, and the
conclusions are the authors' own work. After using this tool, the authors
reviewed and edited the content as needed and take full responsibility for the
content of the publication.

\textbf{Author contributions.} N.E.: conceptualization, methodology, software,
validation, formal analysis, investigation, data curation, visualization,
writing -- original draft. M.E.A.: conceptualization, methodology,
supervision, project administration, writing -- review and editing.

\clearpage
\begin{appendices}

\section{Effects of trimming with mean pooling}\label{app:mean}

\begin{table}[h]
\caption{Effects of trimming (EER percentage points, mean pooling) with 95\%
speaker-level paired bootstrap intervals. Intervals that exclude zero are in
bold. Eval: ASVspoof 2019 LA evaluation set; ITW: In-the-Wild}
\label{tab:effects-mean}
\centering
\footnotesize
\setlength{\tabcolsep}{2pt}
\begin{tabular}{@{}llcccc@{}}
\toprule
Set & L & $\Delta_{\mathrm{test}}$ & $\Delta_{\mathrm{train}}$ &
$\Delta_{\mathrm{total}}$ & $I$ \\
\midrule
Eval &  1 & \textbf{+19.88 [17.43, 22.57]} & \textbf{$-$6.83 [$-$8.75, $-$4.90]} & \textbf{+13.05 [11.30, 14.96]} & \textbf{$-$15.74 [$-$17.75, $-$13.86]} \\
     &  2 & \textbf{+13.27 [11.58, 14.93]} & \textbf{$-$3.82 [$-$5.30, $-$2.16]} & \textbf{+9.45 [8.25, 10.85]} & \textbf{$-$11.74 [$-$13.13, $-$10.16]} \\
     &  3 & \textbf{+10.17 [8.81, 11.63]} & \textbf{$-$2.93 [$-$4.17, $-$1.88]} & \textbf{+7.24 [6.22, 8.39]} & \textbf{$-$8.80 [$-$10.11, $-$7.73]} \\
     &  4 & \textbf{+6.60 [5.71, 7.57]} & \textbf{$-$1.73 [$-$2.63, $-$0.90]} & \textbf{+4.87 [4.03, 5.80]} & \textbf{$-$6.08 [$-$6.98, $-$5.21]} \\
     &  5 & \textbf{+7.81 [6.69, 8.95]} & \textbf{$-$3.98 [$-$4.91, $-$2.99]} & \textbf{+3.84 [3.31, 4.49]} & \textbf{$-$7.70 [$-$8.71, $-$6.72]} \\
     &  6 & \textbf{+8.32 [6.82, 9.81]} & \textbf{$-$5.47 [$-$6.66, $-$4.29]} & \textbf{+2.86 [2.15, 3.51]} & \textbf{$-$7.50 [$-$8.80, $-$6.28]} \\
     &  7 & \textbf{+9.16 [7.27, 10.91]} & \textbf{$-$5.47 [$-$6.87, $-$4.15]} & \textbf{+3.69 [2.79, 4.56]} & \textbf{$-$8.55 [$-$9.88, $-$7.16]} \\
     &  8 & \textbf{+12.78 [10.80, 14.72]} & \textbf{$-$7.50 [$-$9.05, $-$5.93]} & \textbf{+5.28 [4.16, 6.27]} & \textbf{$-$11.60 [$-$13.31, $-$9.80]} \\
     &  9 & \textbf{+13.18 [11.33, 14.84]} & \textbf{$-$7.95 [$-$9.14, $-$6.61]} & \textbf{+5.23 [4.02, 6.29]} & \textbf{$-$12.16 [$-$13.56, $-$10.48]} \\
     & 10 & \textbf{+13.03 [11.10, 14.93]} & \textbf{$-$7.65 [$-$9.06, $-$6.19]} & \textbf{+5.37 [4.22, 6.43]} & \textbf{$-$11.64 [$-$13.24, $-$10.05]} \\
     & 11 & \textbf{+15.32 [13.04, 17.49]} & \textbf{$-$9.80 [$-$11.69, $-$7.84]} & \textbf{+5.52 [4.41, 6.73]} & \textbf{$-$14.69 [$-$16.64, $-$12.59]} \\
     & 12 & \textbf{+14.83 [12.46, 17.17]} & \textbf{$-$8.56 [$-$10.63, $-$6.39]} & \textbf{+6.27 [4.94, 7.63]} & \textbf{$-$14.24 [$-$16.26, $-$12.12]} \\
\midrule
ITW  &  1 & +0.50 [$-$1.83, 2.35] & $-$1.60 [$-$7.98, 6.51] & $-$1.10 [$-$7.43, 6.39] & \textbf{$-$2.12 [$-$3.70, $-$0.12]} \\
     &  2 & $-$0.53 [$-$2.20, 1.06] & $-$3.54 [$-$8.37, 1.56] & $-$4.06 [$-$8.20, 0.64] & \textbf{$-$1.46 [$-$2.76, $-$0.29]} \\
     &  3 & $-$0.31 [$-$1.91, 0.94] & $-$2.90 [$-$6.21, 0.86] & $-$3.21 [$-$6.91, 0.37] & \textbf{$-$3.25 [$-$5.19, $-$1.42]} \\
     &  4 & $-$0.87 [$-$2.33, 0.35] & $-$1.17 [$-$4.13, 2.10] & $-$2.04 [$-$5.07, 0.86] & \textbf{$-$1.32 [$-$2.04, $-$0.47]} \\
     &  5 & $-$0.88 [$-$2.43, 0.50] & \textbf{$-$3.53 [$-$5.90, $-$0.77]} & \textbf{$-$4.41 [$-$6.91, $-$2.18]} & \textbf{$-$2.29 [$-$3.06, $-$1.06]} \\
     &  6 & $-$1.02 [$-$2.71, 0.32] & \textbf{$-$4.86 [$-$6.93, $-$2.19]} & \textbf{$-$5.89 [$-$7.78, $-$3.86]} & $-$0.95 [$-$1.92, 0.41] \\
     &  7 & $-$0.25 [$-$1.52, 0.93] & \textbf{$-$3.07 [$-$5.58, $-$0.12]} & \textbf{$-$3.32 [$-$5.64, $-$0.99]} & \textbf{$-$1.20 [$-$2.19, $-$0.18]} \\
     &  8 & +0.44 [$-$0.42, 1.45] & \textbf{$-$5.41 [$-$8.26, $-$2.59]} & \textbf{$-$4.97 [$-$7.91, $-$2.29]} & 0.00 [$-$0.97, 0.72] \\
     &  9 & +0.26 [$-$0.72, 1.27] & \textbf{$-$4.55 [$-$6.36, $-$2.71]} & \textbf{$-$4.28 [$-$6.28, $-$2.18]} & \textbf{$-$1.09 [$-$1.91, $-$0.05]} \\
     & 10 & +0.48 [$-$1.09, 1.88] & $-$1.65 [$-$4.32, 1.83] & $-$1.17 [$-$3.88, 1.60] & $-$0.75 [$-$1.76, 0.50] \\
     & 11 & $-$0.08 [$-$1.86, 1.42] & +0.29 [$-$2.57, 3.41] & +0.20 [$-$2.34, 2.63] & +0.09 [$-$1.09, 1.49] \\
     & 12 & $-$0.30 [$-$2.04, 1.34] & +0.19 [$-$2.89, 4.28] & $-$0.12 [$-$2.62, 3.24] & +1.03 [$-$0.49, 2.62] \\
\bottomrule
\end{tabular}
\end{table}

\begin{table}[h]
\caption{The two remaining effects of the 2$\times$2 design (EER percentage
points) with 95\% speaker-level paired bootstrap intervals, from the same
resamples as Table~\ref{tab:effects}: training on trimmed audio when the test
audio is clean, $\Delta_{\mathrm{train}}^{\mathrm{C}}$, and
trimming the test audio of the model trained on trimmed audio,
$\Delta_{\mathrm{test}}^{\mathrm{T}}$ (Eqs.~\eqref{eq:dtrainc}--\eqref{eq:dtestt}). Intervals that exclude
zero are in bold}
\label{tab:effects-other}
\centering
\footnotesize
\setlength{\tabcolsep}{2pt}
\begin{tabular}{@{}llcccc@{}}
\toprule
 & & \multicolumn{2}{c}{$\Delta_{\mathrm{train}}^{\mathrm{C}}$} &
 \multicolumn{2}{c}{$\Delta_{\mathrm{test}}^{\mathrm{T}}$} \\
\cmidrule(lr){3-4}\cmidrule(lr){5-6}
Set & L & stats & mean & stats & mean \\
\midrule
Eval &  1 & \textbf{+8.90 [7.43, 10.43]} & \textbf{+8.91 [7.56, 10.56]} & \textbf{+3.92 [2.80, 5.00]} & \textbf{+4.13 [3.09, 5.34]} \\
     &  2 & \textbf{+7.51 [6.36, 8.87]} & \textbf{+7.92 [6.69, 9.31]} & \textbf{+1.60 [0.83, 2.40]} & \textbf{+1.53 [0.78, 2.42]} \\
     &  3 & \textbf{+5.49 [4.55, 6.54]} & \textbf{+5.87 [4.93, 7.02]} & \textbf{+1.56 [0.97, 2.14]} & \textbf{+1.37 [0.73, 1.98]} \\
     &  4 & \textbf{+3.89 [3.14, 4.70]} & \textbf{+4.35 [3.52, 5.18]} & \textbf{+0.62 [0.18, 1.08]} & \textbf{+0.52 [0.03, 1.03]} \\
     &  5 & \textbf{+3.49 [2.85, 4.15]} & \textbf{+3.72 [3.03, 4.48]} & +0.17 [$-$0.36, 0.61] & +0.11 [$-$0.45, 0.72] \\
     &  6 & \textbf{+2.10 [1.59, 2.57]} & \textbf{+2.03 [1.50, 2.49]} & \textbf{+0.79 [0.37, 1.30]} & \textbf{+0.82 [0.41, 1.26]} \\
     &  7 & \textbf{+3.10 [2.57, 3.75]} & \textbf{+3.08 [2.51, 3.68]} & +0.39 [$-$0.42, 1.24] & +0.61 [$-$0.10, 1.29] \\
     &  8 & \textbf{+4.47 [3.79, 5.08]} & \textbf{+4.10 [3.40, 4.76]} & +0.42 [$-$0.57, 1.41] & \textbf{+1.18 [0.11, 2.17]} \\
     &  9 & \textbf{+4.35 [3.56, 5.20]} & \textbf{+4.21 [3.36, 4.94]} & +0.62 [$-$0.36, 1.55] & +1.02 [$-$0.01, 2.07] \\
     & 10 & \textbf{+4.37 [3.58, 5.11]} & \textbf{+3.99 [3.23, 4.69]} & +0.80 [$-$0.21, 1.78] & \textbf{+1.39 [0.31, 2.45]} \\
     & 11 & \textbf{+4.92 [3.83, 6.01]} & \textbf{+4.89 [3.96, 5.85]} & +0.33 [$-$0.45, 1.18] & +0.63 [$-$0.15, 1.40] \\
     & 12 & \textbf{+5.85 [4.77, 6.94]} & \textbf{+5.69 [4.61, 6.69]} & +0.19 [$-$0.73, 1.28] & +0.59 [$-$0.31, 1.70] \\
\midrule
ITW  &  1 & $-$0.26 [$-$6.36, 6.59] & +0.52 [$-$5.90, 7.89] & \textbf{$-$1.39 [$-$2.35, $-$0.56]} & \textbf{$-$1.62 [$-$2.71, $-$0.75]} \\
     &  2 & $-$2.54 [$-$6.80, 2.34] & $-$2.08 [$-$6.55, 3.35] & \textbf{$-$1.93 [$-$2.97, $-$0.79]} & \textbf{$-$1.99 [$-$3.37, $-$0.94]} \\
     &  3 & $-$0.88 [$-$5.42, 4.14] & +0.35 [$-$3.90, 5.24] & \textbf{$-$3.38 [$-$5.53, $-$1.56]} & \textbf{$-$3.56 [$-$5.88, $-$1.63]} \\
     &  4 & $-$1.32 [$-$4.70, 2.09] & +0.14 [$-$2.69, 3.26] & \textbf{$-$1.85 [$-$2.79, $-$0.98]} & \textbf{$-$2.19 [$-$3.23, $-$1.25]} \\
     &  5 & $-$1.41 [$-$3.76, 1.43] & $-$1.24 [$-$3.89, 1.30] & \textbf{$-$2.53 [$-$3.80, $-$1.46]} & \textbf{$-$3.17 [$-$4.47, $-$1.67]} \\
     &  6 & \textbf{$-$3.93 [$-$6.03, $-$1.46]} & \textbf{$-$3.92 [$-$6.03, $-$1.64]} & \textbf{$-$1.60 [$-$2.73, $-$0.64]} & \textbf{$-$1.97 [$-$3.22, $-$0.73]} \\
     &  7 & $-$1.62 [$-$4.34, 1.04] & $-$1.87 [$-$4.55, 1.01] & $-$0.71 [$-$1.46, 0.07] & \textbf{$-$1.45 [$-$2.37, $-$0.53]} \\
     &  8 & \textbf{$-$4.16 [$-$6.86, $-$1.36]} & \textbf{$-$5.41 [$-$8.26, $-$2.69]} & \textbf{+0.95 [0.52, 1.45]} & +0.44 [$-$0.07, 0.80] \\
     &  9 & \textbf{$-$2.53 [$-$4.83, $-$0.37]} & \textbf{$-$3.45 [$-$5.66, $-$1.14]} & $-$0.08 [$-$0.48, 0.48] & \textbf{$-$0.83 [$-$1.68, $-$0.08]} \\
     & 10 & $-$0.67 [$-$3.22, 1.99] & $-$0.90 [$-$3.61, 2.11] & +0.15 [$-$0.31, 0.69] & $-$0.27 [$-$0.96, 0.40] \\
     & 11 & $-$0.43 [$-$3.01, 2.35] & +0.19 [$-$2.36, 2.83] & +0.28 [$-$0.36, 0.87] & +0.01 [$-$0.63, 0.55] \\
     & 12 & $-$0.17 [$-$3.00, 2.97] & $-$0.85 [$-$3.43, 2.50] & \textbf{+0.91 [0.12, 1.73]} & \textbf{+0.73 [0.01, 1.51]} \\

\bottomrule
\end{tabular}
\end{table}

\begin{figure}[t]
\centering
\includegraphics[width=\textwidth]{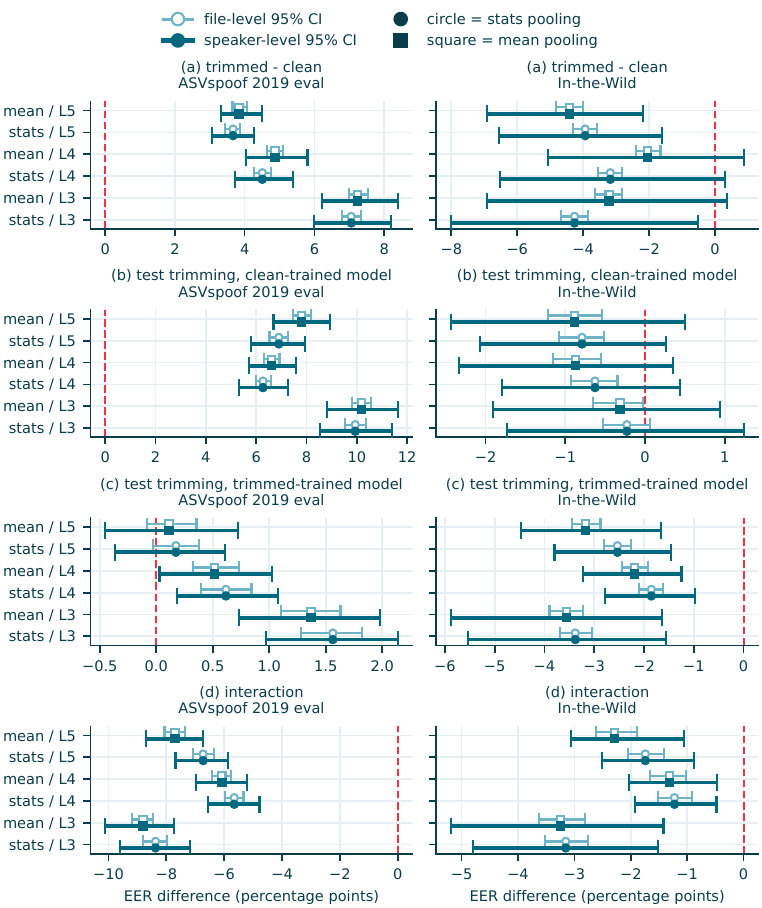}
\caption{Effects of trimming for the six configurations (layers 3--5, mean and
statistics pooling) on the evaluation set (left) and on In-the-Wild (right):
(a) trimmed minus clean ($\Delta_{\mathrm{total}}$), (b) trimming the test
audio of the model trained on clean audio ($\Delta_{\mathrm{test}}$), (c)
trimming the test audio of the model trained on trimmed audio
($\Delta_{\mathrm{test}}^{\mathrm{T}}$), and (d) the
interaction $I$. Circles: statistics pooling; squares: mean pooling; the red
dashed line marks zero. All intervals
are paired, class-stratified percentile bootstrap CIs (1{,}000 resamples) of
the seed-mean EER (EER per seed, averaged over five seeds); light: file
level, dark: speaker level}
\label{fig:forest}
\end{figure}

\clearpage
\section{Interval construction for the rank agreement}\label{app:bca}
\setcounter{table}{0}

\begin{table}[h]
\caption{Kendall's $\tau$ between the clean and trimmed layer rankings with
95\% speaker-level percentile and BCa intervals from the same 1{,}000
resamples; $z_0$: bias-correction constant (ties counted one half); $a$:
acceleration (leave-one-speaker-out jackknife within class); ties: share of
resamples equal to the observed $\tau$; $p$: exact two-sided $p$-value for
$n = 12$}
\label{tab:bca}
\centering
\footnotesize
\begin{tabular}{@{}lllcccrrrr@{}}
\toprule
Threshold & Set & Pooling & $\tau$ & Percentile & BCa & $z_0$ & $a$ & Ties & $p$ \\
\midrule
40~dB & Eval & stats & 0.58 & [0.39, 0.64] & [0.52, 0.70]  & 0.86    & 0.019    & 18\% & 0.009 \\
      & Eval & mean  & 0.58 & [0.42, 0.64] & [0.52, 0.67]  & 0.89    & $-$0.007 & 15\% & 0.009 \\
      & ITW  & stats & 0.73 & [0.42, 0.85] & [0.58, 0.93]  & 0.49    & 0.017    & 10\% & 0.0005 \\
      & ITW  & mean  & 0.76 & [0.39, 0.82] & [0.67, 0.91]  & 0.99    & 0.027    & 8\%  & 0.0002 \\
\midrule
20~dB & Eval & stats & 0.48 & [0.27, 0.52] & [0.45, 0.55]  & 1.21    & 0.010    & 10\% & 0.031 \\
      & Eval & mean  & 0.39 & [0.27, 0.48] & [0.27, 0.48]  & $-$0.09 & $-$0.015 & 23\% & 0.086 \\
      & ITW  & stats & 0.52 & [0.27, 0.82] & [0.01, 0.70]  & $-$0.46 & $-$0.025 & 10\% & 0.021 \\
      & ITW  & mean  & 0.58 & [0.27, 0.79] & [0.36, 0.83]  & 0.24    & $-$0.003 & 11\% & 0.009 \\
\midrule
15~dB & Eval & stats & 0.42 & [0.27, 0.52] & [0.30, 0.55]  & 0.22    & $-$0.012 & 19\% & 0.063 \\
      & Eval & mean  & 0.36 & [0.27, 0.48] & [0.21, 0.45]  & $-$0.39 & $-$0.016 & 21\% & 0.116 \\
      & ITW  & stats & 0.64 & [0.27, 0.79] & [0.48, 0.93]  & 0.71    & 0.024    & 6\%  & 0.003 \\
      & ITW  & mean  & 0.48 & [0.24, 0.76] & [0.18, 0.70]  & $-$0.23 & $-$0.016 & 9\%  & 0.031 \\
\bottomrule
\end{tabular}
\end{table}

\end{appendices}

\clearpage
\bibliography{sn-bibliography}

\end{document}